\documentclass[fleqn,usenatbib]{mnras}

\usepackage{newtxtext,newtxmath}

\usepackage[T1]{fontenc}

\DeclareRobustCommand{\VAN}[3]{#2}
\let\VANthebibliography\thebibliography
\def\thebibliography{\DeclareRobustCommand{\VAN}[3]{##3}\VANthebibliography}

\usepackage{graphicx}	
\usepackage{amsmath}	
\usepackage{xcolor}
\usepackage{multirow}
\usepackage{booktabs}

\newcommand{\todo}[1]{\textcolor{black}{#1}}

\title[Precision Abundances in NGC~288 \& NGC~362 I]{Peeking beneath the precision floor I: metallicity spreads and multiple elemental dispersions in the globular clusters NGC~288 and NGC~362}

\author[S. Monty et al.]{Stephanie Monty$^{1,2,3}$\thanks{E-mail: sm2744@cam.ac.uk},
David Yong$^{2,3}$,
Anna F. Marino$^{4,5}$,
Amanda I. Karakas$^{6,3}$,
Madeleine McKenzie$^{2,3}$,
\newauthor
Frank Grundahl$^{7}$,
Aldo Mura-Guzm{\'a}n$^{2,3}$,
\\
$^{1}$ Institute of Astronomy, University of Cambridge, Madingley Rd, Cambridge, CB3 0HA, UK\\
$^{2}$ Research School of Astronomy and Astrophysics, Mt Stromlo Observatory, Weston Creek, ACT 2611, Australia\\
$^{3}$ ARC Centre of Excellence for All Sky Astrophysics in 3 Dimensions (ASTRO 3D), Australia\\
$^{4}$ Instituto Nazionale di Astrofisica - Osservatorio Astronomico di Padova, Vicolo dell'Osservatorio 5, IT-35122,  Padua, Italy\\
$^{5}$ Instituto Nazionale di Astrofisica - Osservatorio Astrofisico di Arcetri, Largo Enrico Fermi, 5, IT-50125, Firenze, Italy\\
$^{6}$ School of Physics \& Astronomy, Monash University, Clayton, VIC 3800, Australia\\
$^{7}$  Stellar Astrophysics Centre, Department of Physics and Astronomy, Aarhus University, DK-8000 Aarhus C, Denmark
}

\date{Accepted XXX. Received YYY; in original form ZZZ}

\pubyear{2022}

\begin{document}
\label{firstpage}
\pagerange{\pageref{firstpage}--\pageref{lastpage}}
\maketitle

\begin{abstract}
The view of globular clusters (GCs) as simple systems continues to unravel, revealing complex objects hosting multiple chemical peculiarities. Using differential abundance analysis, we probe the chemistry of the Type I GC, NGC~288 and the Type II GC, NGC~362 at the 2\% level for the first time. We measure 20 elements and find differential measurement uncertainties on the order 0.01-0.02~dex in both clusters. The smallest uncertainties are measured for \ion{Fe}{I} in both clusters, with an average uncertainty of $\sim$0.013~dex. Dispersion in the abundances of Na, Al, \ion{Ti}{I}, Ni, \ion{Fe}{I}, Y, Zr, Ba and Nd are recovered in NGC~288, none of which can be explained by a spread in He. This is the first time, to our knowledge, a statistically significant spread in $s$-process elements and a potential spread in metallicity has been detected in NGC~288. In NGC~362, we find significant dispersion in the same elements as NGC~288, with the addition of Co, Cu, Zn, Sr, La, Ce, and Eu. Two distinct groups are recovered in NGC~362, separated by 0.3~dex in average differential $s$-process abundances. Given strong correlations between Al and several $s$-process elements, and a significant correlation between Mg and Si, we propose that the $s$-process rich group is younger. This agrees with asymptotic giant branch star (AGB) enrichment between generations, if there is overlap between low- and intermediate-mass AGBs. In our scenario, the older population is dominated by the $r$-process with a $\Delta^{\mathrm{La}}-\Delta^{\mathrm{Eu}}$ ratio of $-0.16\pm0.06$. We propose that the $r$-process dominance and dispersion found in NGC~362 are primordial. \end{abstract}

\begin{keywords}
\todo{techniques: spectroscopic -- stars: abundances -- globular clusters: general -- globular clusters: individual: NGC 288 -- globular clusters: individual: NGC 362 -- stars: Population II}
\end{keywords}



\section{Introduction}
\label{sec:intro}
Globular clusters are among the most well-studied enigmas in modern astrophysics. Once prized as theoretical test beds for stellar evolution, nucleosynthesis and stellar dynamics, the assumption of their simplicity has continued to unravel in recent years \citep[a selection of reviews on the subject include;][]{gratton12, bastian18, gratton19, milone22}. Chemically, globular clusters (GCs) are anything but simple, with many showing star-to-star abundance variations involving light elements (e.g., O and Na) \citep{yong09, carretta14, carretta15, yong15} as well as spreads in iron \citep{gratton12b, yong14, yong16, marino18} and other heavy elements \citep{gratton12b, yong14}.

Early detection of the chemical complexity of GCs, like a large spread in strength of the cyanogen molecule (CN) \citep{freeman75, norris75, bessel76, cottrell81}, has since been shown to be indicative of multiple stellar populations (MSPs) formed through distinct episodes of star formation. The chemical differences between generations are attributed to high temperature hydrogen burning involving the elements C, N, O, F, Na, Mg and Al in sites including asymptotic giant branch stars (AGBs), fast rotating massive stars, binaries and supermassive stars \citep[see][for a review]{gratton04}. The use of narrow band filters aboard the Hubble Space Telescope (HST) in the last decade has been groundbreaking in revealing the seemingly ubiquitous appearance of MSPs in Milky Way GCs \citep{marino08, piottohugs, milone17}.

Despite the discovery that almost all GCs are not simple stellar systems, some GCs remain outliers even in this increasingly complex space. Cataloguing the characteristics of MW GCs has revealed two populations, termed  ``Type I'' and ``Type II''. Grouping GCs into these two types was done using pioneering high resolution photometry and chemical abundance analysis by \cite{milone17} and \cite{marino19}. The majority ($\sim80\%$) of GCs are of Type I and display two distinct populations in chemical and colour-magnitude space due to differences in O, Na, N and He abundances (the result of H-burning, as mentioned). Type I GCs are also characterised by homogeneous abundances in elements heavier than Si. Type II GCs account for the other $\sim20\%$ of MW GCs and are characterised by additional complexity in their HST pseudo two-colour diagrams (``chromosome maps'') \citep{milone17}. Many Type II GCs demonstrate multiple populations beyond just the two found in Type I GCs, often displaying spreads in metallicity and/or slow neutron capture elements \citep{marino11c, marino15, marino21}.

One of the most famous Type II GCs is $\omega$ Centauri ($\omega$-Cen), the most most massive and chemically complex GC \citep{marino11b, marino12}. Owing to both of these characteristics, it has been proposed to be the nucleus of an accreted dwarf galaxy (dGal) \citep{majewski00, bekkiomega}. Another Type II GC, M~54, is known to be the nucleus of the presently disrupting Sagittarius dGal \citep{ibata94, ibata95}. To explain the anomalous chemistry of Type II GCs, an extra-galactic origin has been proposed for these GCs, be-it as the nuclei of accreted dGals or as a member of a GC system of an accreted dGal \citep{bekki12, marino15, dacosta16, marino19}. To investigate this further, \cite{milone20} included information on the GCs dynamics and found that seven out of 13 (7/13) Type II GCs likely share a common origin (accreted as part of one event). \todo{One of the more famous Type II GCs that is \textit{not} classified as accreted, is the massive GC M~22. Chemically, M~22 has often been compared to $\omega$-Cen given its large dispersion in metallicity \citep[recently confirmed using differential abundance analysis,][]{mckenzie22} and heavy elements \citep{marino11c} and yet it is firmly connected to the MW disc - making accretion unlikely.}

Among the least massive Type II GCs is NGC~362, which displays both a spread in metallicity ($\sim0.12$~dex), detectable with low resolution spectroscopy \citep{husser20}, and a Ba-enhanced population of stars occupying a second red giant branch (RGB) \citep{carretta13}. While a spread in slow neutron capture elements is likely, thus far no obvious spread in rapid neutron capture elements (namely Eu) has been detected \citep{worley10}. NGC~362 is also known to harbour at least two stellar populations and is unique among MSP-hosting GCs given that the older generation of stars is located in the central regions of the cluster \citep{lim16}. Despite its classification as a Type II, and unlike its more massive counterparts, NGC~362 is not considered a likely candidate to be the nucleus of an accreted dGal \citep{pfeffer21}. Although, an extra-galactic origin for this GC could still be likely. We investigate this possibility in an upcoming companion paper.

The common companion to NGC~362 in the literature is the Type I GC NGC~288, which together with NGC~362, forms the canonical second parameter problem pair. The second parameter problem manifests as the appearance of distinctly different horizontal giant branch morphologies in the colour magnitude diagrams (CMDs) of nearly identical metallicity GCs. This is highlighted using the purple bounding boxes in Fig.~\ref{fig:cmds} for NGC~288 (left) and NGC~362 (right). Chemically, in the study of \citet{shetrone00} NGC~288 was found to be slightly more metal-poor than NGC~362 ([Fe/H]$=-1.39$ vs [Fe/H]$=-1.33$) and slightly more enhanced in Al, Na and Ba. Like most GCs, NGC~288 is also known to host two populations of stars but without any clear difference in metallicity between the two (at $R\sim18,000$), as is expected for a Type I GC \citep{hsyu14}. 
 
The primary aim of this study is to re-examine the chemical abundances within the two GCs, NGC~288 and NGC~362, at the ~0.01 dex (2\%) precision level. To do this we use the technique of differential abundance analysis to remove as many systematic sources of error as possible. Such measurements should thereby, (i) provide new insight into the chemical homogeneity of each cluster and, (ii) reveal any unexpected elemental correlations which could be indicative of the cluster formation environment and/or internal evolution. 

The paper is organised as follows. Sec.~\ref{sec:obsanalysis} describes the observational dataset and analysis technique. Sec.~\ref{sec:chemabund_clusthist} presents the recovered dispersion in each element, in the context of the element individually and as a member of a nucleosynthetic group. Correlations within each group are also explored in this section. Sec.~\ref{sec:unexpeccorr} explores the unexpected correlations between elements not found in the same nucleosynthetic group. The unexpected correlations are then discussed in the context of cluster formation and evolution. Finally, Sec.~\ref{sec:sumconc} provides a summary of the major results and the conclusions of the paper.

\begin{table*}
	\centering
	\caption{Target information for the fifteen stars selected for re-analysis with VLT/UVES. Gaia IDs \citep{gaiadr2, gaiaedr3} and membership probability as taken from \citet{vasilievbaumgardt} are listed in the first two columns. The total exposure time resulting from summing $N$ exposures is listed prior to the date the observation was collected, which is then followed by the total number of exposures. $V$-band magnitudes are taken from \citet{shetrone00} Table 1, references listed therein.}
	\label{tab:obs}
	\begin{tabular}{lllllllll} 
		\hline
		Star & Gaia ID & Mem. Prob. & R.A. & Decl. & $V$  & Exp. Time & Obs. Date & $N$ Im. \\
		    & & & [J2000]   & [J2000]   &  & [s]   & & \\
		\hline
		NGC288-20c & 2342903118077555840 & 1.0 & 00:52:43.30 & -26:36:57.09 & 12.96 & 6000.0 & July 15, 2005 & 2 \\
		... & ... & ... & ... & ... & ... & ... & July 21, 2005 & ...\\
		NGC288-281 & 2342904488170510592 & 0.99 & 00:52:58.46 & -26:36:06.12 & 13.27 & 9000.0 & Aug. 18, 2005 & 3\\
		NGC288-287 & 2342907584843612416 & 1.0 & 00:52:46.67 & -26:35:08.10 & 14.72 & 15000.0 & Aug. 24, 2005 & 5\\
		... & ... & ... & ... & ... & ... & ... & Aug. 25, 2005 & ...\\
		... & ... & ... & ... & ... & ... & ... & Sept. 11, 2005 & ...\\
		NGC288-338 & 2342904763048460416 & 1.0 & 00:52:52.80 & -26:34:38.73 & 13.63 & 6000.0 & Sept. 15, 2005 & 2\\
		NGC288-344 & 2342904732985400704 & 1.0 & 00:52:52.88 & -26:35:20.09 & 13.27 & 9000.0 & Aug. 18, 2005 & 3\\
		... & ... & ... & ... & ... & ... & ... & Sept. 11, 2005 & ...\\
		NGC288-351  & 2342904659969201024 & 1.0 & 00:52:52.51 & -26:36:04.03 & 13.54 & 9000.0 & Aug. 14, 2005 & 3\\
		\hline
		NGC362-1137 & 4690886864638749312 & 1.0 & 01:02:59.23 & -70:49:43.8 & 13.02 & 9000.0 & July 21, 2005 & 3\\
		... & ... & ... & ... & ... & ... & ... & Aug. 5, 2005 & ... \\
		... & ... & ... & ... & ... & ... & ... & Aug. 20, 2005 & ... \\
		NGC362-1334 & 4690839448199896704 & 1.0 & 01:03:38.19 & -70:52:00.6 & 12.77 & 9000.0 & Aug. 24, 2005 & 3\\
		... & ... & ... & ... & ...  & ... & ... & Aug. 26, 2005 & ...\\
		... & ... & ... & ... & ...  & ... & ... & Sept. 11, 2005 & ...\\
		NGC362-2127 & 4690839448199896704 & 1.0 & 01:02:37.12 & -70:50:33.0 & 12.95 & 9000.0 & July 20, 2005 & 3 \\
		NGC362-1401 & 4690839723077800320 & 0.99 & 01:03:36.08 & -70:50:50.9 &  12.63 & 6000.0 & Aug. 23, 2005 & 2\\
		NGC362-1423 & 4690839791797273216 & 1.0 & 01:03:33.48 & -70:49:35.0 & 12.77 & 6000.0 & Aug. 23, 2005 & 2\\
		NGC362-1441 & 4690886795919334912 & 1.0 & 01:03:22.52 & -70:48:38.7 & 12.72 & 6000.0 & Aug. 22, 2005 & 2\\
		NGC362-77 & 4690886727199867904 & 1.0 & 01:03:25.04 & -70:49:56.2 & 12.72 & 9000.0 & Aug. 23, 2005 & 3\\
		... & ... & ... & ... & ... & ... & ... & Aug. 24, 2005 & ...\\
		NGC362-MB2 & 4690886658480344320 & 1.0 & 01:03:07.53 & -70:49:43.7 & 12.94 & 6000.0 & Aug. 22, 2005 & 2\\
		\hline
	\end{tabular}
\end{table*}

\section{Observations \& Analysis}
\label{sec:obsanalysis}
\subsection{Target Selection}
\label{sec:targets}

\begin{figure}
	\includegraphics[width=\columnwidth]{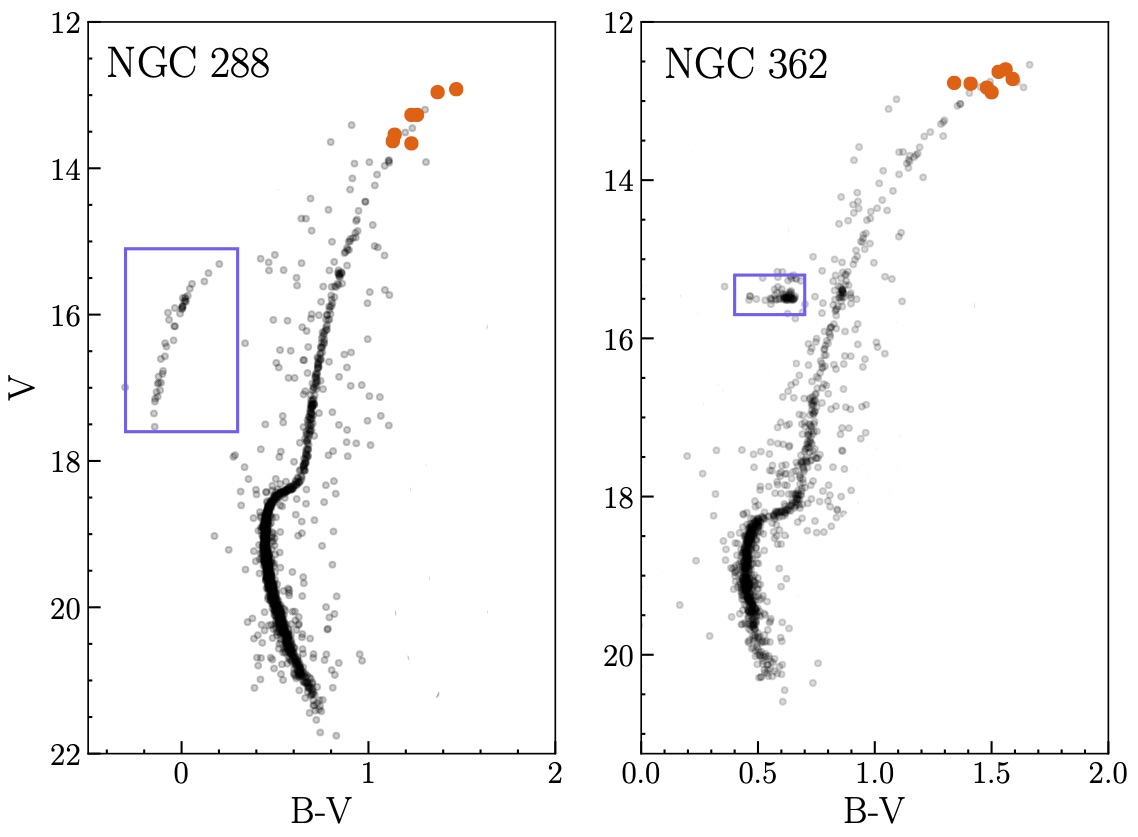}
    \caption{Colour-magnitude diagrams for the clusters, NGC~288 (\textit{left}) and NGC~362 (\textit{right}) created using cleaned catalogues from \citet{pbscat}. The different horizontal giant branch morphology of the two clusters (characteristic of the second parameter problem) is highlighted by the purple bounding box in both clusters. The stars chosen for this study are highlighted in orange in both clusters.}
    \label{fig:cmds}
\end{figure}

\begin{figure}
	\includegraphics[width=\columnwidth]{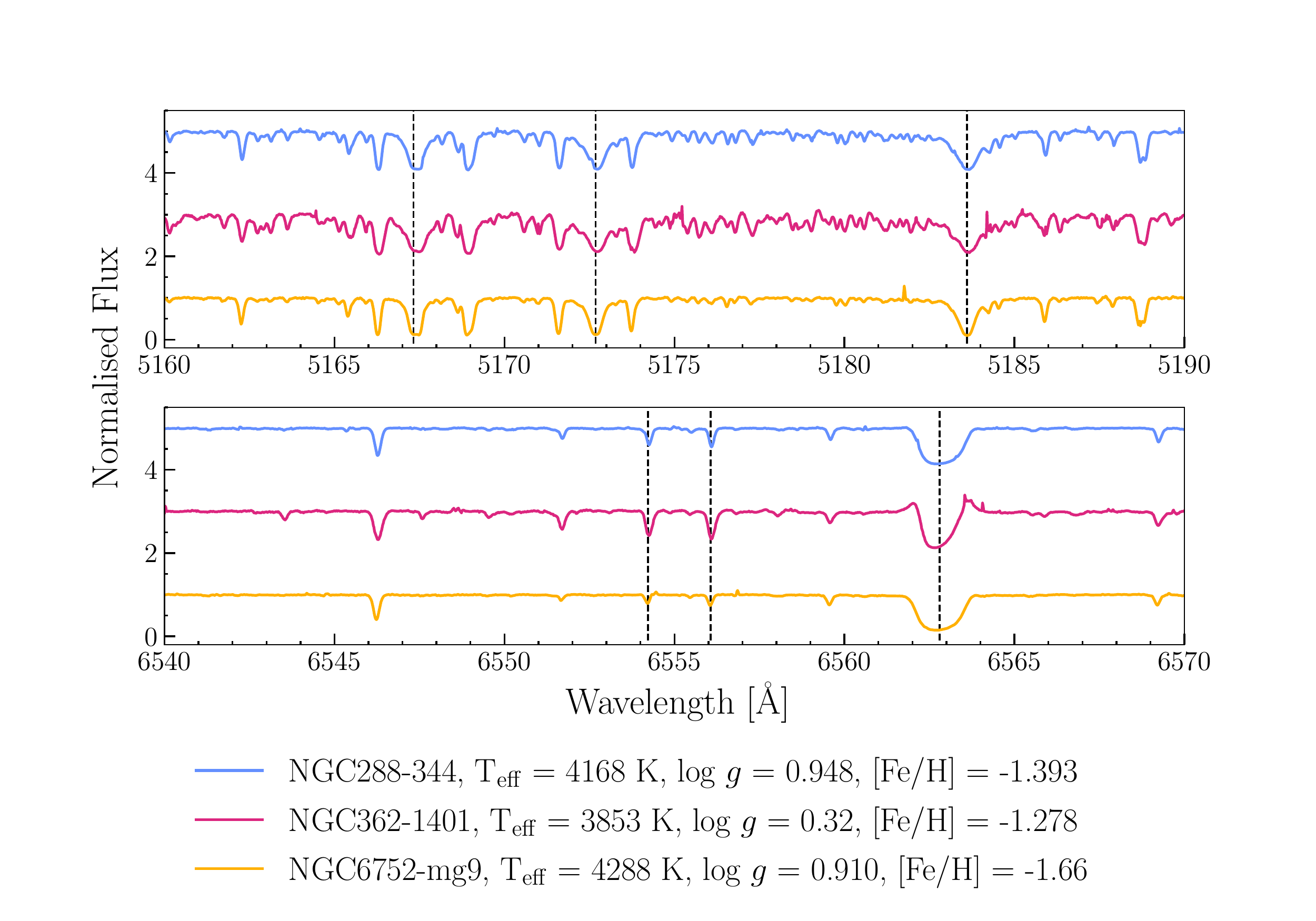}
    \caption{Example spectra showing two regions of the VLT/UVES spectra for a sample star from each cluster and the reference star mg9. In the top panel the Mgb lines are shown at 5167.3~\AA, 5172.7~\AA\, and 5183.6~\AA\, via the vertical dashed lines. The bottom panel highlights two \ion{Ti}{I} lines at $\sim6554$~\AA\, and $\sim6556$~\AA\,, alongside $\mathrm{H}\alpha$ at $\sim6562$~\AA. The best-fit stellar parameters recovered via the method described in Section~\ref{sec:stellarparams} are listed for the program stars. The stellar parameters for mg9 are from taken from \citet{yong2013}.}
    \label{fig:examplespectra}
\end{figure}

We select a total of 14 stars from the original work of \cite{shetrone00} to reanalyse, six in NGC~288 and eight in NGC~362. These stars are shown in orange in Fig.~\ref{fig:cmds} highlighting their locations near the tip of the red giant branch (RGB)\footnote{Although a fascinating subject, we want to avoid investigating the phenomena of MSPs in our clusters. Cross-matching with the catalogues of \citet{piottohugs}, resulted in HST UV photometry for only two of our NGC~362 stars and three of our NGC~288 stars. Stars in both clusters appeared to occupy the same region of each cluster's chromosome map.}. As differential abundance analysis requires both high signal-to-noise (S/N) and high resolution spectra, we select our sample from amongst the brightest stars in the original study and re-observe them using VLT/UVES \citep[R$\sim110,000$,][]{uves}. Example spectra for two of the program stars, NGC288-344 and NGC362-1401 are shown in Fig.~\ref{fig:examplespectra} highlighting the Mgb lines at $\sim5100$~\AA\, in the top panel and two \ion{Ti}{I} lines at $\sim6554$~\AA\, and $\sim6556$~\AA\, in the bottom panel, alongside $\mathrm{H}\alpha$ at $\sim6562$~\AA\,.

The choice of stars in our study is also governed by the requirement that they act as approximate ``stellar siblings'', meaning they span a very small range in stellar parameters. The original stellar parameters derived by \citet{shetrone00} were used in the initial selection to ensure this. Note that we will refer to the effective temperature ($\mathrm{T}_{\mathrm{eff}}$), surface gravity (log~$g$) and metallicity (expressed as [Fe/H]) as the fundamental stellar parameters in this work. Observational information and characteristics of the stellar spectra are presented in Table~\ref{tab:obs}, alongside the Gaia DR2/DR3 \citep{gaiadr2, gaiaedr3} IDs and the membership probability as determined by \citet{vasilievbaumgardt} using proper motions from Gaia EDR3 \citep{gaiaedr3}.

\subsection{Line List \& Equivalent Width Measurements}
\label{sec:ews}
The line list to measure abundances was created by combining the line lists from the studies of \cite{yong2013}, \cite{battaglia17} and \cite{ji2020streams} (and references therein). In the case of overlapping lines, priority was given to the more recent publication. Two additional species were added to the final line list, \ion{Zr}{II} \citep{roederer18} and \ion{Sr}{I} from NIST \citep{nistsr}. Hyperfine structure (HFS) corrections were applied to the 5853.67\AA, 6141.71\AA\, and 6496.9\AA\, lines of \ion{Ba}{II}, the 5303.53\AA\, line of \ion{La}{II}\, and the 6645.10\AA\, line of \ion{Eu}{II} using \texttt{linemake}\footnote{\url{https://github.com/vmplacco/linemake}} \citep{lalinemake, eulinemake, linemake}. \todo{To apply the HFS corrections, the additional transitions output from \texttt{linemake} were added to the input linelist.} 

Initial equivalent width (EW) measurements were made using the automated EW measurement software, \texttt{DAOSpec} \citep{daospec}. Secondary measurements were also made using \texttt{REvIEW}\footnote{\url{https://github.com/madeleine-mckenzie/REvIEW}} a \textsc{python}-based automated tool for EW measurements described in \citet{mckenzie22}. Initial cuts were made to only include lines with EW measurements in the range [5, 100]~m\AA\ as measured by \texttt{DAOSpec}. The two overlapping measurement sets were then compared for every line in common. To identify poor measurements in either method, lines were initially flagged if the standard deviation of the two measurements was greater than 5~m\AA. This proved a more conservative method than culling by using an arbitrary value of sigma. 

Following initial flag assignments, the flagged lines were then examined by hand using the \texttt{splot} routine within \textsc{IRAF}\footnote{IRAF is distributed by the National Optical Astronomy Observatory, which is operated by the Association of Universities for Research in Astronomy (AURA) under cooperative agreement with the National Science Foundation}. The hand-measured value most often lay between the two automated measurement values and thus the mean value of the two was assigned as the final EW. In the case that the values differed greatly from the hand-measured value, the hand-measured value was adopted. Finally, additional lines were added for elements with exclusively larger than 100~m\AA\ measurements (e.g. \ion{Ba}{II}, \ion{Mn}{I}, \ion{V}{I}.)

A sample of EWs for several stars in the study, including the reference star \citep[mg9, or B3169][]{yong2013, buonanno86} discussed in the upcoming section, are given in Table~\ref{tab:ews}. The full version of Table~\ref{tab:ews} is included with the online supplementary material.

\subsection{\label{sec:stellarparams}Stellar Parameter Determination}
Stellar parameters and abundances were determined using the \textsc{python} tool \texttt{q2} \citep{q2} in a two step process. Using \texttt{q2} to communicate to the 1D local thermodynamic equilibrium (LTE) radiative transfer code \texttt{MOOG} \citep{moog} and a set of $\alpha$-enhanced MARCS model atmospheres \citep{marcs}, initial stellar parameters were found using the classical spectroscopic approach in a differential sense with respect to the reference star \citep{melendez09}. We do not consider departures from LTE in this study as the range of stellar parameters spanned by our program stars is small (a result of our choice of ``stellar siblings''). We select the same reference star as \citet{yong2013} to perform our differential analysis, namely the star mg9 found near the tip of the RGB in NGC~6752. This choice was motivated by the similarities in stellar parameters between mg9 and our program stars \todo{(T$_{\mathrm{eff}}$=4288~K, log~$g$=0.91, $\xi=1.72$~km/s, [Fe/H]=-1.66)}, and to place the abundances on the same scale as the study of \citet{yong2013}.

The initial values of effective temperature (T$_{\mathrm{eff}}$) and microturbulence ($\xi$) were found via minimising the slopes of $\Delta^{\mathrm{FeI}}$ versus excitation potential and $\Delta^{\mathrm{FeI}}$ versus log~(EW/wavelength) respectively. The differential abundance of \ion{Fe}{I} ($\Delta^{\mathrm{FeI}}$)  is determined line-by-line as $\delta A_{\mathrm{line}}=A_{\mathrm{line}}^{\mathrm{program\,star}}-A_{\mathrm{line}}^{\mathrm{mg9}}$ where $A$ is the abundance measurement associated with each line. An initial value of surface gravity (log~$g$) was found via imposing ionization equilibrium between the $\Delta^{\mathrm{FeI}}$ and $\Delta^{\mathrm{FeII}}$ abundances. Note that $\Delta^{\mathrm{X}}$ refers to the differential elemental abundance relative to the references star mg9 (here and throughout), and hence all excitation balances and ionization equilibria were achieved in a differential sense. Determining stellar parameters in this manner has been shown to provide accurate results \citep{nissen18}.

As a starting point for the minimisation process, the stellar parameters from \cite{shetrone00} were fed to \texttt{q2}. Preliminary step sizes of $\pm200$~K, $\pm0.5$~cm/s$^{2}$ and $\pm0.5$km/s in T$_{\mathrm{eff}}$, log~$g$ and micro-turbulence respectively were then selected for the initial exploration. \texttt{q2} uses an iterative process to converge on the best-fit stellar parameters by finding an initial minimum, re-sampling the atmospheric grid with a smaller step size surrounding the initial solution and then re-determining the best-fit stellar parameters. This process is continued until an absolute minimum is found, yielding the final stellar parameters. 

Using the initial stellar parameters determined via the process described above, preliminary \ion{Fe}{I} and \ion{Fe}{II} abundances were found in both the absolute and differential sense. In the case of the reference star mg9, abundances were only determined in the absolute sense. This is described in more detail in the next section. The initial \ion{Fe}{I} and \ion{Fe}{II} abundances were then plotted as a function of wavelength to examine outliers. Initially, a $1.5\sigma$ cull was performed to remove the bulk of the outliers. These lines were then visually examined to identify blended lines, poor continuum placement and/or inaccurate measurement of EW. Based on this they were remeasured, or removed from the line list for that star entirely. 

Following the culling procedure, the second step in the process was performed by re-determining the stellar parameters via running the 1.5$\sigma$-culled linelist through \texttt{q2}. The initial stellar parameters determined during the first round of minimisation were used as the starting point for \texttt{q2} and the step sizes were reduced to $\pm30$~K, $\pm0.1$~cm/s$^{2}$ and $\pm0.1$km/s in T$_{\mathrm{eff}}$, log~$g$ and micro-turbulence respectively. Formal error analysis was performed by \texttt{q2} in a purely differential sense taking into account co-variances and following the approach of \citet{epstein10}. The final stellar parameters adopted for the remainder of the study are listed in Table~\ref{tab:stellarparams}, alongside the stellar parameters of the reference star. 

Comparing our final parameters to \citet{shetrone00}, we find average absolute differences of $\Delta$T$_{\mathrm{eff}}=30\pm21$~K (NGC~288) and $\Delta$T$_{\mathrm{eff}}=46\pm28$~K (NGC~362) in effective temperature, $\Delta$log~$g=0.16\pm0.10$~dex (NGC~288) and $\Delta$log~$g=0.19\pm0.10$~dex (NGC~362) in surface gravity, $\Delta\xi=0.12\pm0.0.10$~km/s (NGC~288) and $\Delta\xi=0.21\pm0.16$~km/s (NGC~362) in microturbulence and finally, comparing the metallicity of our adopted atmospheric models to those of \citet{shetrone00}, yields average differences of $\Delta$[Fe/H]$_{\mathrm{model}}=0.05\pm0.03$~dex (NGC~288) and $\Delta$[Fe/H]$_{\mathrm{model}}=0.07\pm0.05$~dex (NGC~362).

\begin{table}
	\centering
	\caption{Final stellar parameters for the program stars in NGC~288 and NGC~362 respectively (separated by the third horizontal line) derived using the process outlined in Section~\ref{sec:stellarparams} and adopted for the remainder of the study. \todo{The \textit{differential} uncertainties on the stellar parameters are also listed.} The stellar parameters for the NGC~6752 reference star, mg9, taken from \citet{yong2013} are also listed.}
	\label{tab:stellarparams}
	\begin{tabular}{lllll} 
		\hline
		Star & T$_{\mathrm{eff}}$ & log~$g$ & $\xi$ & [Fe/H]\\
		    & [K] & [cm/s$^{2}$]   & [km/s]   &  \\
		\hline
		20c & $4109\pm9$ & $0.76\pm0.05$ & $1.78\pm0.03$ & $-1.376\pm0.01$ \\
		281 & $4144\pm7$ & $0.86\pm0.06$ & $1.61\pm0.02$ & $-1.362\pm0.01$ \\
		287 & $4335\pm12$ & $1.18\pm0.05$ & $1.65\pm0.02$ & $-1.438\pm0.01$ \\
		338 & $4314\pm14$ & $1.28\pm0.06$ & $1.60\pm0.03$ & $-1.390\pm0.01$ \\
		344 & $4168\pm5$ & $0.95\pm0.06$ & $1.67\pm0.03$ & $-1.393\pm0.01$ \\
		351 & $4264\pm16$ & $1.13\pm0.05$ & $1.69\pm0.03$ & $-1.422\pm0.02$ \\
		403 & $3977\pm20$ & $0.55\pm0.05$ & $1.62\pm0.03$ & $-1.326\pm0.02$ \\\hline
		1137 & $4071\pm20$ & $0.62\pm0.07$ & $1.76\pm0.05$ & $-1.243\pm0.02$ \\
		1334 & $4043\pm23$ & $0.62\pm0.07$ & $1.83\pm0.06$ & $-1.175\pm0.03$ \\
		2127 & $4124\pm14$ & $0.59\pm0.05$ & $1.85\pm0.04$ & $-1.271\pm0.02$ \\
	    1401 & $3853\pm17$ & $0.32\pm0.14$ & $2.00\pm0.10$ & $-1.278\pm0.04$ \\
		1423 & $4046\pm15$ & $0.27\pm0.09$ & $2.14\pm0.07$ & $-1.301\pm0.02$ \\
		1441 & $3942\pm11$ & $0.41\pm0.10$ & $1.91\pm0.07$ & $-1.179\pm0.02$ \\
		77 & $4127\pm17$ & $0.50\pm0.07$ & $1.97\pm0.06$ & $-1.297\pm0.02$ \\
		MB2 & $4085\pm12$ & $0.40\pm0.10$ & $2.59\pm0.12$ & $-1.328\pm0.02$ \\\hline
		mg9 & 4288 & 0.91 & 1.72 & -1.66 \\
		\hline
	\end{tabular}
\end{table}

\subsection{Differential Abundances}
\label{sec:difabund}
The stellar abundances for both clusters were determined on a line-by-line basis, in a strictly differential sense, relative to the reference star mg9. The process of relative abundance determination is described in detail in \citet{melendez12} and \cite{yong2013}. Briefly, relative abundances are determined by measuring the line-by-line \textit{difference} in abundance between the program star and reference star. The total average differential abundance for every element is then $\Delta^{\mathrm{X}}=\frac{1}{N}\sum_{i=1}^{N}\delta A_{\mathrm{line}}$. The final differential abundances for our program stars are given in Tables~\ref{tab:ngc288_abund} and \ref{tab:ngc362_abund}. The absolute abundances for the reference star mg9 are also listed. Note that no corrections for departures from LTE (non-LTE, NLTE) have been applied to any of the differential abundances. As mentioned in the previous section, NLTE effects are minimised in this study through selection of ``stellar siblings''. Any remaining \textit{differential} NLTE effects are considered negligible. \todo{HFS and isotopic splitting corrections were made by running \texttt{MOOG} in \texttt{blend} mode within \texttt{q2}.} Errors associated with the differential abundances are calculated as in \citet{q2}.
                                            
\section{Chemical Abundance Dispersion \& Expected Elemental Correlations}
\label{sec:chemabund_clusthist}
In this section we present the chemical abundance results for the two clusters, focusing on the dispersion in each element and correlations among expected elements. Each element is grouped by nucleosynthetic source as given by \citet{burbidge} and includes a discussion of both individual elements and the group as a whole. Explanations for the recovered elemental dispersions will also be discussed in this section in terms of each cluster individually. Implications for the two cluster accretion scenarios are discussed in forthcoming sections and in our upcoming companion paper.

\subsection{Statistically Significant Spreads in Abundance}
\label{sec:abundspreads}
Abundances of several elements in \textit{both} clusters show star-to-star variations and potentially significant dispersion. Here and throughout, we define dispersion in an element as the standard deviation in the set of abundance measurements for stars in that cluster. To explore the likelihood that these spreads are real, we simulated the expected dispersion due to measurement errors alone in several well-measured elements ($N$ lines > 3) and compared this to the recovered dispersion. To simulate the dispersion due to uncertainties, we created a Gaussian probability distribution centred around the average abundance measurement for each element, in each cluster. 

\begin{figure}
	\includegraphics[scale=0.5]{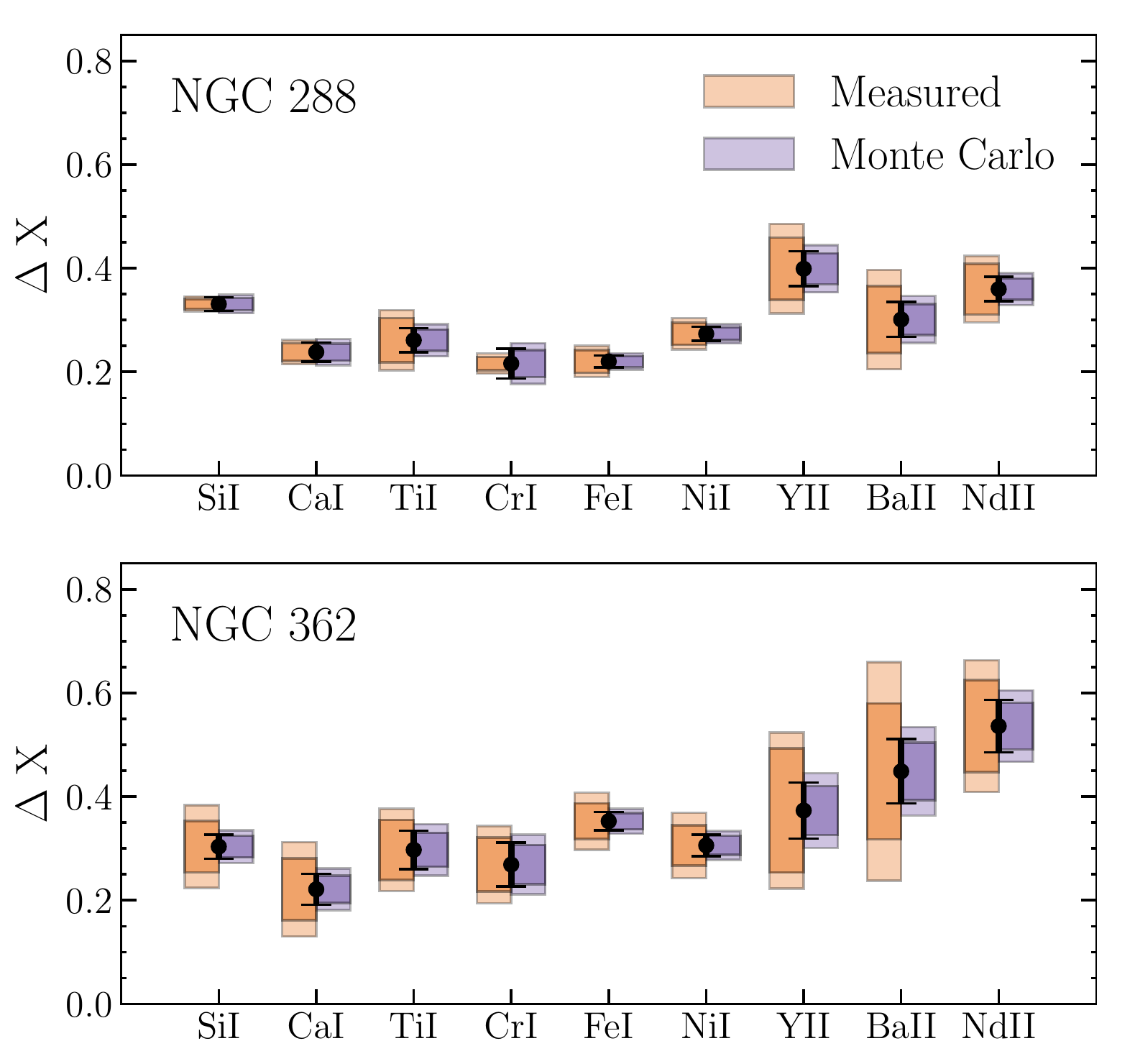}
    \caption{Comparison between simulated (purple) and measured (orange) elemental dispersion for elements with three or more lines measured for NGC~288 (top) and NGC~362 (bottom). In both cases the lighter shade shows the amplitude of the spread, while the darker shade shows the standard deviation. The $s$-process enhanced star NGC362-1441 is excluded from the calculations of dispersion.}
    \label{fig:bothdisp}
\end{figure}

The width of the distribution was dictated by the average measurement error, again for each element, in each cluster. Several thousand Montecarlo realisations were made, drawing six stars (in the case of NGC~288) and seven stars (in the case of NGC~362) to simulate the final dispersion. Only seven stars were selected in NGC~362 because star NGC362-1441 was excluded from the investigation (the reason behind this choice is discussed in upcoming sections). The results of this investigation are shown in Fig.~\ref{fig:bothdisp}, where the measured spreads are shown in orange centred around the average measured value, and the simulated spreads are shown in purple. The black points and error bars represent the average dispersion measured and the measurement uncertainty. The amplitude of each spread is shown using the lighter shade and the standard deviation is shown using the darker shade. 

In NGC~362 all spreads presented in Fig.~\ref{fig:bothdisp} are real, in that they cannot be reproduced through measurement errors alone. In NGC~288, the heavy elements Y, Ba and Nd show the most obvious spread. However, a small but significant spread in Fe is also recovered. We assess individual element dispersions and further discuss what we consider to be genuine elemental spreads in the next section.

\begin{table*}
    \centering
    \caption{Average (differential) cluster abundances in the elements discussed in Sec.~\ref{sec:abundspreads} ($\Delta^{\mathrm{X}}_{\mathrm{ave}}$), along with the average measurement error in each element ($\sigma_{\mathrm{meas,ave}}$, the contribution from He is not included), the dispersion in the element within the cluster ($\sigma\Delta^{\mathrm{X}}$) and the uncertainty on the dispersion ($\sigma\Delta^{\mathrm{X}}_{\mathrm{error}}$) as defined in Sec.~\ref{sec:globexpl}. Two measurements sets are given for NGC~362, the first is without including the $s$-process enhanced star 1441 (no $s$-rich) and second, including 1441 ($s$-rich).}
   	\label{tab:all_disp}
    \begin{tabular}{l|lll|lll|lll|lll}
	\hline
    & & \multirow{1}{*}{NGC~288} & & & & & & \multirow{1}{*}{NGC~362} & & & & \\
    & & & & & \multirow{1}{*}{no $s$-rich} & & & & \multirow{1}{*}{$s$-rich} & & & \\
    Element & $\Delta^{\mathrm{X}}_{\mathrm{ave}}$ & $\sigma_{\mathrm{meas,ave}}$ & $\sigma\Delta^{\mathrm{X}}$ & $\sigma\Delta^{\mathrm{X}}_{\mathrm{error}}$ & $\Delta^{\mathrm{X}}_{\mathrm{ave}}$ & $\sigma_{\mathrm{meas,ave}}$ & $\sigma\Delta^{\mathrm{X}}$ & $\sigma\Delta^{\mathrm{X}}_{\mathrm{error}}$ & $\Delta^{\mathrm{X}}_{\mathrm{ave}}$ & $\sigma_{\mathrm{meas,ave}}$ & $\sigma\Delta^{\mathrm{X}}$ & $\sigma\Delta^{\mathrm{X}}_{\mathrm{error}}$ \\
    \hline
    \ion{Na}{I} & 0.282 & 0.023 & 0.252 & 0.080 & 0.217 & 0.033 & 0.213 & 0.062 & 0.276 & 0.032 & 0.241 & 0.064 \\
    \ion{Al}{I} & 0.031 & 0.013 & 0.109 & 0.034 & -0.013 & 0.023 & 0.191 & 0.055 & 0.011 & 0.022 & 0.189 & 0.051 \\
    \ion{Mg}{I} & 0.223 & 0.117 & 0.060 & 0.019 & 0.247 & 0.077 & 0.037 & 0.011 & 0.253 & 0.078 & 0.040 & 0.011 \\
    \ion{Si}{I} & 0.317 & 0.019 & 0.012 & 0.004 & 0.297 & 0.028 & 0.049 & 0.014 & 0.299 & 0.029 & 0.046 & 0.012 \\
    \ion{Ca}{I} & 0.236 & 0.021 & 0.017 & 0.005 & 0.218 & 0.033 & 0.059 & 0.017 & 0.234 & 0.033 & 0.064 & 0.017 \\
    \ion{Ti}{I} & 0.266 & 0.025 & 0.041 & 0.013 & 0.289 & 0.040 & 0.058 & 0.017 & 0.301 & 0.039 & 0.060 & 0.016 \\
    \ion{Ti}{II} & 0.222 & 0.027 & 0.026 & 0.008 & 0.229 & 0.037 & 0.082 & 0.024 & 0.238 & 0.037 & 0.080 & 0.021 \\
    \ion{Cr}{I} & 0.215 & 0.030 & 0.013 & 0.004 & 0.272 & 0.045 & 0.052 & 0.015 & 0.283 & 0.045 & 0.056 & 0.015\\
    \ion{Cr}{II} & 0.279 & 0.067 & 0.094 & 0.030 & 0.337 & 0.046 & 0.100 & 0.029 & 0.481 & 0.043 & 0.245 & 0.065 \\
    \ion{Fe}{I} & 0.224 & 0.015 & 0.021 & 0.007 & 0.353 & 0.023 & 0.035 & 0.010 & 0.361 & 0.023 & 0.039 & 0.010 \\
    \ion{Fe}{II} & 0.240 & 0.031 & 0.017 & 0.005 & 0.335 & 0.054 & 0.041 & 0.012 & 0.347 & 0.053 & 0.046 & 0.012 \\
    \ion{Co}{I} & 0.257 & 0.015 & 0.031 & 0.010 & 0.322 & 0.023 & 0.062 & 0.018 & 0.330 & 0.024 & 0.061 & 0.016 \\
    \ion{Ni}{I} & 0.275 & 0.017 & 0.021 & 0.007 & 0.302 & 0.026 & 0.039 & 0.011 & 0.310 & 0.026 & 0.042 & 0.011 \\
    \ion{Cu}{I} & 0.661 & 0.032 & 0.156 & 0.049 & 0.549 & 0.076 & 0.137 & 0.040 & 0.615 & 0.078 & 0.266 & 0.071 \\
    \ion{Zn}{I} & 0.107 & 0.018 & 0.088 & 0.028 & 0.164 & 0.035 & 0.073 & 0.021 & 0.193 & 0.035 & 0.100 & 0.027 \\
    \ion{Sr}{I} & 0.117 & 0.018 & 0.035 & 0.011 & 0.094 & 0.035 & 0.235 & 0.068 & 0.123 & 0.032 & 0.230 & 0.062 \\
    \ion{Y}{II} & 0.391 & 0.035 & 0.059 & 0.019 & 0.361 & 0.056 & 0.120 & 0.035 & 0.396 & 0.054 & 0.157 & 0.042 \\
    \ion{Zr}{II} & 0.352 & 0.022 & 0.033 & 0.010 & 0.314 & 0.032 & 0.150 & 0.043 & 0.345 & 0.028 & 0.162 & 0.043 \\
    \ion{Ba}{II} & 0.303 & 0.035 & 0.064 & 0.020 & 0.448 & 0.064 & 0.131 & 0.038 & 0.548 & 0.059 & 0.213 & 0.057 \\
    \ion{La}{II} & 0.365 & 0.047 & 0.036 & 0.011 & 0.419 & 0.046 & 0.078 & 0.022 & 0.460 & 0.044 & 0.146 & 0.039 \\
    \ion{Ce}{II} & 0.368 & 0.041 & 0.038 & 0.012 & 0.379 & 0.048 & 0.056 & 0.016 & 0.417 & 0.046 & 0.145 & 0.039 \\
    \ion{Nd}{II} & 0.359 & 0.026 & 0.049 & 0.015 & 0.521 & 0.053 & 0.089 & 0.026 & 0.559 & 0.051 & 0.127 & 0.034\\
    \ion{Eu}{II} & 0.402 & 0.024 & 0.018 & 0.006 & 0.543 & 0.037 & 0.090 & 0.026 & 0.559 & 0.034 & 0.093 & 0.025 \\
    \hline
\end{tabular}
\end{table*}

\subsubsection{Global Explanations for Observed Spreads}
\label{sec:globexpl}
Statistically significant dispersion among the majority of well-measured elements was also recovered by \citet{yong2013} in the disc-like GC NGC~6752. In their study, \citet{yong2013} discussed four possible scenarios to explain both the recovered dispersion and positive correlations among unexpected elements. The first two were i) systematic errors in the determination of the stellar parameters and ii) unaccounted for star-to-star CNO abundance variations. Both were ruled out by \citet{yong2013} through i) persistence of the correlations under variation of stellar parameters and ii) a change in the CNO abundances in the stellar atmospheres from ``CN-weak''  to ``CN-strong'' resulting in a statistical strengthening of existing correlations. Instead, \citet{yong2013} found two more likely explanations for the elemental dispersions and strong positive correlations, i) \textit{ab-initio} He variations between the stars and ii) inhomogenous chemical evolution in the proto-cluster environment.

In this section we focus on the first potential explanation, He abundance variations, to explain the dispersions we observe and discuss the second explanation later in the context of unexpected elemental correlations. As discussed in \citet{yong2013}, He-abundance variations would manifest in two ways. The first being a change to the model atmospheres, as they assume a fixed He-abundance, resulting in a change in stellar parameters. The second being a change to the global metal mass fraction ($Z$), and thus all values of [X/H], as a result of a change in the He mass fraction ($Y$).

A recent determination of the He spread between first and second generation stars in our two clusters predicts small He spreads in both \citep{milone18}. In NGC~362, the average spread is expected to be $\Delta~Y=0.008\pm0.006$ and in NGC~288 the average spread is slightly higher at $\Delta~Y=0.01\pm0.010$. The maximum He differences in each clusters are $\Delta~Y_{\mathrm{max}}=0.026\pm0.008$ and $\Delta~Y_{\mathrm{max}}=0.016\pm0.012$ in NGC~362 and NGC~288 respectively. Using equation 12 in \citet{stromgren82}, and adopting the \texttt{MARCS} model atmosphere He abundance ($Y=0.25$) as our starting value, we can determine the average and maximal shift in log~$g$ due to the He spreads in both GCs. Note that this is under the assumption that He-enriched atmospheres behave like He-normal atmospheres with higher values of log~$g$ in the case of both F dwarfs and metal poor giants \citep{stromgren82, lind11}. 

In the case of NGC~362, where the average log~$g$ value is 0.52 among the program stars, the maximum difference in log~$g$ is $\sim0.015$. For NGC~288, where the average log~$g$ value is 0.94, the maximum difference is $\sim0.017$. Using the average cluster He abundance spreads the changes to log~$g$ are even smaller, on the order $0.01$ in NGC~288 and $0.005$ in NGC~362. As the average and maximum changes to log~$g$ due to He abundance spreads are so small and well within the uncertainties attached to our determinations of log~$g$ in both clusters, we neglect this effect. 

The second effect we investigate is the global offset to [X/H], under the assumption of a fixed metal fraction in both clusters, and the condition that the total mass fraction is equal to unity ($Z+X+Y=1$). When adopting a solar metal fraction of $Z_{\odot}=0.0139$ \citep{asplund20}, fixing the metal fraction in each cluster and exploring the maximum change to the He fraction, we find a maximum change in [X/H] of $\sim0.015$~dex in NGC~362 and $\sim0.009$~dex in NGC~288. In the following sections we include a discussion of the existence of the recovered dispersion for each element in the presence of these He-induced offsets. 

In the last decade, an ab-initio spread in He abundance has emerged as a strong contender to solve the second parameter problem (as introduced in Sec.~\ref{sec:intro}) \citep{milone14}. The other contender is likely age \citep{dotter10}. Given the small differences in [X/H] predicted by a He-spread in our two clusters and the primary goals of this study, we do not investigate the second parameter problem in these two clusters beyond this.

Dispersion ($\sigma_{\Delta^{\mathrm{[X/H]}}}$) values for individual elements are shown in Figs.~\ref{fig:lightelemdisp}, \ref{fig:fepeakelemdisp} and \ref{fig:heavyelemdisp} for the two GCs under study as well as the MW ``disc'' GC, NGC~6752 using the data from \cite{yong2013}. As in Sec.~\ref{sec:abundspreads}, the star NGC362-1441 is excluded from the determination of dispersion. The dispersion values are also listed for both clusters in Table~\ref{tab:all_disp}, alongside the average cluster abundances and errors. In the case of NGC~362, we include the averages determined both with and without the $s$-process enhanced star 1441. Each dispersion shown in Fig.s~\ref{fig:lightelemdisp}-\ref{fig:heavyelemdisp} has an associated uncertainty, defined as $\frac{\sigma_{\mathrm{[\Delta X/Fe]}}}{\sqrt{2N - 2}}$ where $N$ is the number of measured abundances ($\sim$ the number of stars in each cluster). These uncertainties are listed in Table~\ref{tab:all_disp}, under the $\sigma\Delta^{\mathrm{X}}_{\mathrm{error}}$ column headers. Arrows in Fig.s~\ref{fig:lightelemdisp}-\ref{fig:heavyelemdisp} denote the dispersion measured for elements with single line measurements. Finally, the quadrature sum of the average measurement error and the contribution from a spread in He for each cluster, in each element, is shown alongside each dispersion value. 

\subsubsection{Light Elements}
\label{sec:lighelemintro}
Before starting our discussion, we must first define what we consider to be a ``genuine'' abundance spread in a particular element. Throughout this section, we will define an abundance spread to be genuine if the dispersion in the element exceeds 1.5 times the quadrature sum of the average measurement error and the maximum spread introduced by an ab-intio spread in He ($\sigma_{\mathrm{ave+He}}$). 

\begin{figure}
    \centering
    	\includegraphics[width=\linewidth]{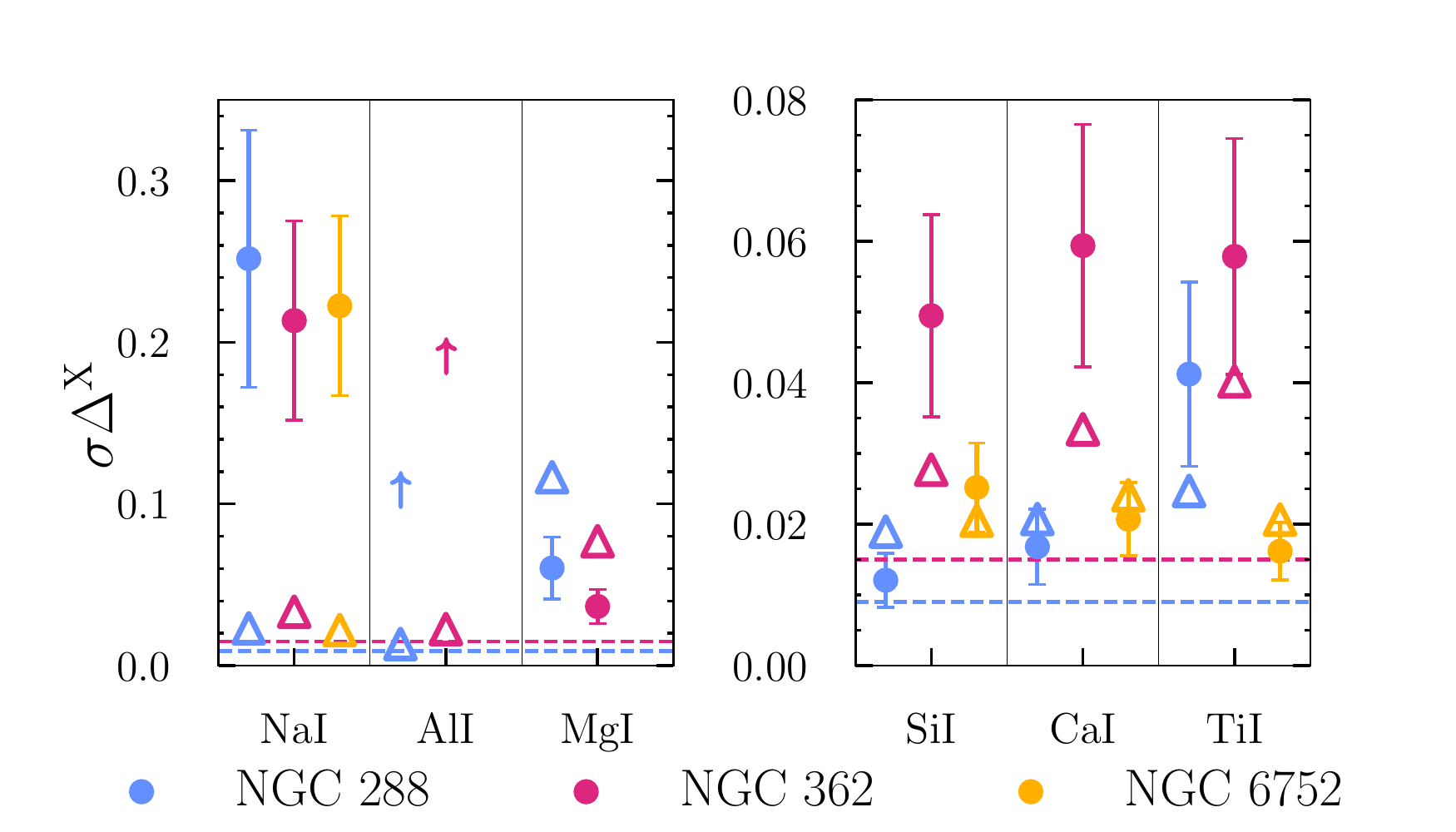}
        \caption{Intrinsic dispersion vs. the quadrature sum of the average measurement error and predicted spread in He for the light elements and the intermediate element Ti in NGC~288, and NGC~362 and the disc-like GC, NGC~6752. The uncertainty on the dispersion is calculated as described in Section~\ref{sec:abundspreads}. The triangle markers denote the quadrature sum of the average measurement and the spread in He for each GC \citep[He contribution $\approx0.018$ in NGC~6752,][]{yong2013}. The contribution from an ab-initio spread in He alone is marked with a dashed line for each GC (at 0.015 for NGC~362 and 0.009~dex for NGC~288). Single-line measurements are marked using arrows.}
        \label{fig:lightelemdisp}
\end{figure}

Starting with the light elements Na, Al, Mg and the intermediate elements Si, Ca and Ti, we display the recovered dispersion values in Fig.~\ref{fig:lightelemdisp} alongside $1.5\sigma_{\mathrm{ave+He}}$. Nucleosynthetically, the odd-elements Na and Al are primarily synthesised via hydrostatic C-burning in massive stars and released into the ISM via core-collapse supernovae (CCSNe) \citep{understell}. They are also produced in both massive and AGB stars through the NeNa and MgAl cycles, where Ne and Mg are converted into Na and Al respectively \citep{karakas10, nomoto13}.

Historically, Mg, Si, Ca and Ti, have been classified as members of the $\alpha$-element family which are synthesised in massive stars and predominantly released via CCSNe  \citep{kobayashi06}. Type Ia SNe also contribute an appreciable amount of Si, Ca and Ti to the ISM, although not in quantities comparable to the production of Fe and Fe-peak elements (to be discussed in the next section) \citep{chiaki20}. This contribution from Type Ia may cause both $\alpha$ and Fe-peak elements to follow similar trends.
 
Examining Fig.~\ref{fig:lightelemdisp}, typical light element spreads are recovered for the two GCs (in \ion{Na}{I} and \ion{Al}{I}) and considered to be genuine, in agreement with what was found in NGC~6752 \citep{yong2013} and the majority of MW GCs \citep[in fact this is often the basis for defining a GC, see][and references therein]{gratton19}. Calculating the dispersion in Na and Al using our definition and the abundance measurements in \citet{shetrone00} for stars in common, our recovery of a dispersion in Na and Al is supported by the results of \citet{shetrone00}. In NGC~288 their abundance measurements yield dispersion values of $\sigma_{\mathrm{Na}}=0.28\pm0.09$ and $\sigma_{\mathrm{Al}}=0.18\pm0.06$, and in NGC~362 (neglecting star 1441), their measurements yield $\sigma_{\mathrm{Na}}=0.22\pm0.06$ and $\sigma_{\mathrm{Al}}=0.25\pm0.07$, in good agreement with what is shown in Table~\ref{tab:all_disp}. 

No detectable spread in Mg is found in either of the clusters, although there are relatively large error bars associated with the measurements (likely due to line saturation). Spreads in the remaining light and $\alpha$-elements Si and Ca differ between the two clusters under study, with NGC~362 displaying genuine spreads in both Si and Ca and the scale of the dispersions in NGC~288 appearing more similar to NGC~6752. This is also apparent in \ion{Ti}{I}, although to a lesser extent, where both cluster dispersions are considered genuine. Note that the larger uncertainty associated with the Ti measurements is likely due to the larger number of measured lines leading to increased scatter in this case. The lack of any significant spread in Ca in NGC~288 is in agreement with previous observations of narrow-band Ca photometry of the cluster \citep{lim15}. 

When only considering Ca, Si, \ion{Ti}{I} and \ion{Ti}{II}, the weighted average alpha element dispersion values ($\Delta^{\alpha}-\Delta^{\mathrm{Fe}}$, analogous to [$\alpha$/Fe]) in the two clusters are $\sigma_{\Delta^{\alpha\mathrm{/Fe}}}=0.02\pm0.006$ for NGC~288 and $\sigma_{\Delta^{\alpha\mathrm{/Fe}}}=0.05\pm0.02$ for NGC~362. The uncertainties on the dispersion values are determined as described in Sec.~\ref{sec:globexpl}. The average value of $\sigma_{\mathrm{ave+He}}$ for the four $\alpha$-elements is 0.03~dex in NGC~288 and 0.04~dex in NGC~362. In the case of NGC~362, the average $\alpha$-element dispersion value is slightly larger than the uncertainty introduced by $\sigma_{\mathrm{ave+He}}$ ($\sim1.25\times$ larger). In the case of NGC~288, the spread can be explained by measurement errors and a He-spread alone. If the dispersion in NGC~362 is genuine, this is likely the first detection of a spread which has been predicted \citep{marino18} but previously undetected \citep{kovalev}.

To evaluate the scale of our $\alpha$-element dispersion values, we can compare our values to the two lightest Type II GCs (both more massive than NGC~288), NGC~1261 (M$=1.82\times10^{5}$~M$_{\odot}$) and NGC~6934 (M=$1.36\times10^{5}$~M$_{\odot}$) \citep{baumgardthilker}. Note that both are similar in metallicity to our two GCs (NGC~1261 [Fe/H]$\sim-1.3$, NGC~6934 [Fe/H]$\sim-1.6$ \citet{marino21}). Using the most recent abundance measurements from \citet{marino21} and the same $\alpha$-elements, we find a dispersion of $\sigma_{[\alpha\mathrm{/Fe}]}=0.03\pm0.004$ in NGC~1261 and $\sigma_{[\alpha\mathrm{/Fe}]}=0.04\pm0.003$ in NGC~6934. Although the average measurement errors are $\sim0.10$ dex in both clusters, making it difficult to conclude that the dispersion is real. 

If the $\alpha$-dispersion is to be believed in the two clusters (which may not be the case for NGC~288), our values place NGC~362 well within the low-mass range of Type II GCs and NGC~288 not far outside the range - despite being a Type I GC. In total we find five out of six (5/6, in Na, Al, Si, Ca and Ti) genuine dispersion measurements in the light and $\alpha$-elements in NGC~362 and three out of size (3/6, in Na, Al and Ti) in NGC~288. Of these dispersion measurements, the largest and smallest dispersions are $\sigma_{\Delta^{\mathrm{Na}}}\sim0.25\pm0.08$ and $\sigma_{\Delta^{\mathrm{Si}}}\sim0.01\pm004$ in NGC~288 and $\sigma_{\Delta^{\mathrm{Na}}}\sim0.21\pm0.06$ and $\sigma_{\Delta^{\mathrm{Mg}}}\sim0.04\pm0.11$ in NGC~362, respectively.

Another interesting thing to note is the opposite direction of the trend in $\alpha$-element dispersion found within GCs vs. dGals. Massive GCs display larger spreads in both light elements and $\alpha$-elements \citep[e.g. $\omega$-Cen with a spread of 0.2~dex in Si,][]{johsonomega}, while massive dGals show small spreads in $\alpha$-elements compared to their less massive counterparts. For example, in the LMC \citep[total mass, $\mathrm{M}_{\mathrm{T}}=1.4\times10^{11}$~M$_{\odot}$, ][]{erkel19} $\sigma_{\mathrm{[Si/Fe]}}\sim0.10$ over $\Delta\mathrm{[Fe/H]}=0.2$~dex \citep{pompeia08, berg15}, while in the less massive dGal, Sculptor \citep[total mass, $\mathrm{M}_{\mathrm{T}}=3.4\times10^{8}$~M$_{\odot}$, ][]{battaglia08}, $\sigma_{\mathrm{[Si/Fe]}}\sim0.3$ over $\Delta\mathrm{[Fe/H]}=0.3$~dex \citep{hill19}. Note that the studies of \citet{pompeia08} and \citet{hill19} were both completed using the VLT/FLAMES spectrograph with comparable resolutions and S/N ($\sim80$). This trend in dGals has been proposed to be due to inhomogeneous mixing in low gas-mass environments and has been supported by simulations \citep{revaz12}. Beyond the light element variations, perhaps the ``heavy'' $\alpha$-element dispersion in GCs may also be distinct from dGals.

\subsubsection{Fe-Peak Elements \& Spreads in Metallicity}
\label{sec:fepeak_disp}
\begin{figure}
    \centering
    	\includegraphics[width=\linewidth]{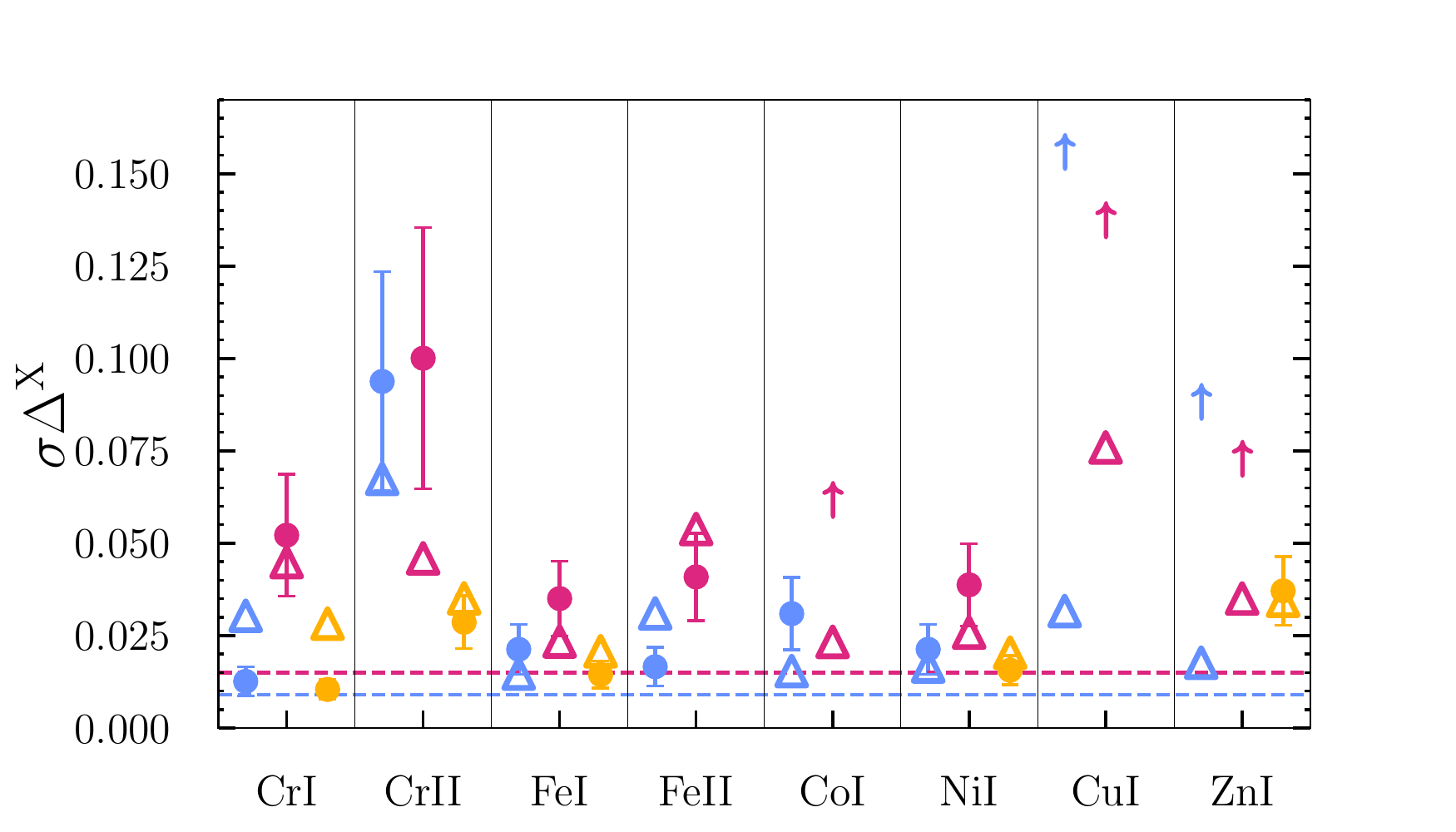}
        \caption{Same as Fig.~\ref{fig:lightelemdisp} for the Fe-peak elements. Implications are discussed in Sec.~\ref{sec:fepeak_disp}.}
        \label{fig:fepeakelemdisp}
\end{figure}

Fig.~\ref{fig:fepeakelemdisp} explores the spread in the iron(Fe)-peak elements, Cr, Co, Ni, Cu, Zn and Fe itself (both ionisation states). Fe-peak elements are primarily synthesised through the capture of $\alpha$-particles (He-nuclei) onto lighter nuclei like Si. This follows the photo-distintegration of heavy nuclei which release the excess $\alpha$-particles \citep{understell}. Both processes occur via nuclear statistical equilibrium during supernovae explosions \citep{woosley02}. Enrichment of Cr, Fe and Ni is dominated by Type Ia SNe, while Co, Cu, Zn are predominantly released in hypernovae (HNe), ultra-energetic CCSNe with progenitor masses greater than 20 M$_{\odot}$ and explosion energies 10 times greater than that of regular CCSNe \citep{chiaki20}.

The Fe-peak elements show a mixture of detectable and non-detectable spreads in abundance \todo{in both clusters.} In the case of both NGC~288 and 362, we infer the existence of genuine spreads in \ion{Fe}{I}, \ion{Ni}{I}, \ion{Co}{I} and \ion{Cu}{I} in the two GCs. Note that the Co and Cu abundances are derived from single line measurements \todo{and thus the uncertainties could be underestimated as they only account for the uncertainties on the stellar parameters.} The spreads in \ion{Fe}{I} and \ion{Ni}{I}, in great agreement in both clusters, are perhaps the most interesting as they are the best-measured elements in our study. On average between 20-33 \ion{Ni}{I} lines and 67-103 \ion{Fe}{I} lines were measured in the two GCs, with NGC~288 having more measureable lines in general due to the higher S/N of the data. The dispersion in Ni and Fe is $\sigma_{\Delta^{\mathrm{Ni, Fe}}}=0.021\pm0.007$, $\sigma_{\Delta^{\mathrm{Ni}}}=0.039\pm0.011$ and $\sigma_{\Delta^{\mathrm{Fe}}}=0.035\pm0.010$ in NGC~288 and NGC~362 respectively.

In both GCs, based on our definition, we \todo{suggest the} dispersion in both Ni and \ion{Fe}{I} to be genuine in NGC~362 and \textit{potentially} genuine in NGC~288, as the dispersion values are only $1.2-1.4$~$\times$~$\sigma_{\mathrm{ave+He}}$, with the spread in \ion{Fe}{I} being more statistically significant. The detection of a spread in Fe in NGC~362 is at odds with the recent findings of \citet{vargas22}, who declare that no spread is detectable after using three different measurement techniques and considering their average measurement error (on the order 0.05~dex, compared to 0.023~dex in our study). However, the scale of the Fe-dispersion detected in our study is in good agreement with all three dispersion measurements made in \citet{vargas22} and given the average uncertainty associated with our differential \ion{Fe}{I} abundances, we maintain our classification of the spread as genuine. 

\begin{figure}
    \centering
    	\includegraphics[width=\linewidth]{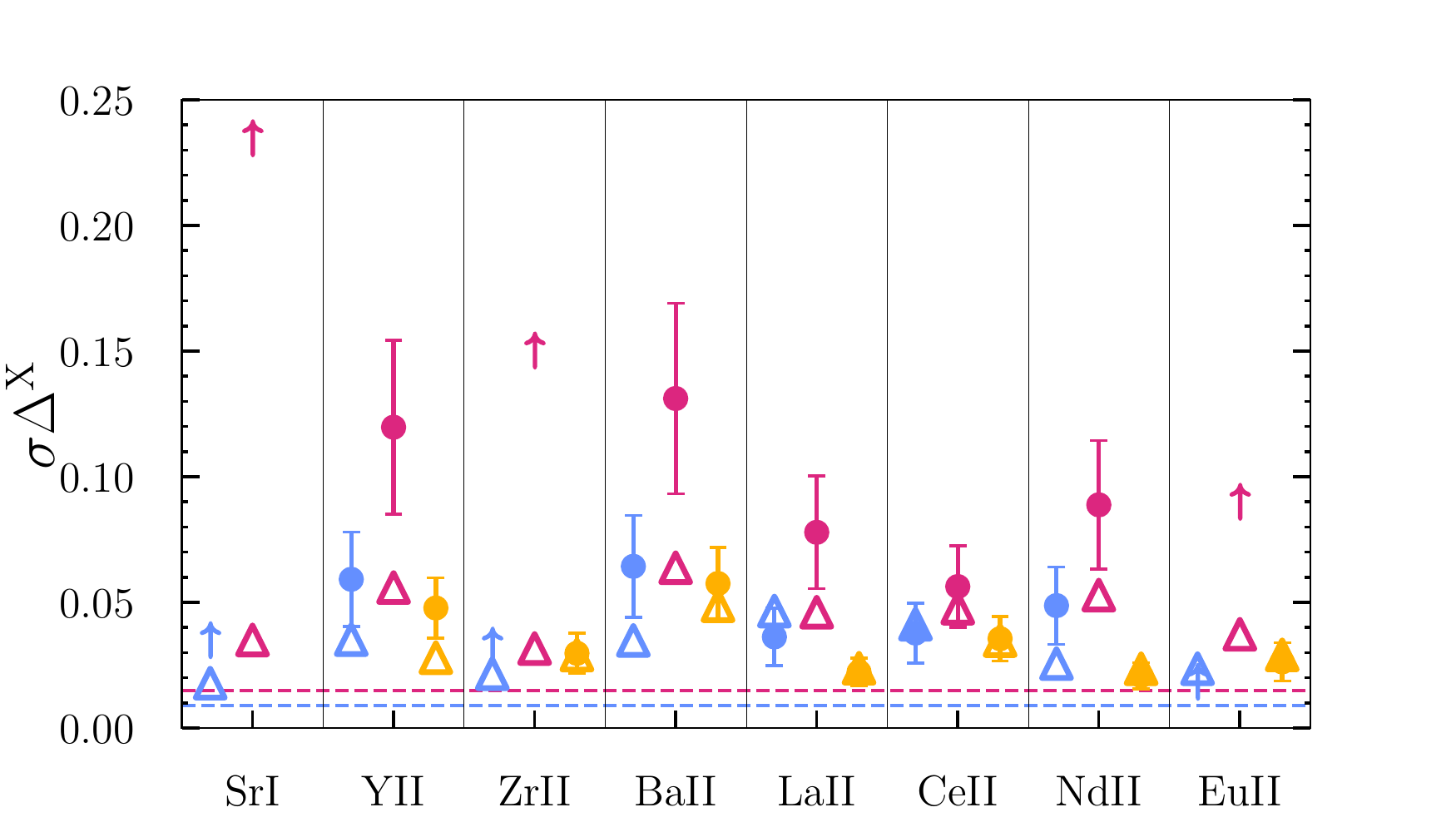}
        \caption{Same as Fig.~\ref{fig:lightelemdisp} for the heavy $s$-process and $r$-process elements discussed in Sec.~\ref{sec:heavyelemdisp}.}
        \label{fig:heavyelemdisp}
\end{figure}

Marginal spreads in both Ni and \ion{Fe}{I} were also detected in NGC~6752 \citep{yong2013}. To the best of our knowledge, this is the first time a statistically significant spread in metallicity has been observed in NGC~288 \citep{carrmetspread}. In total we find six out of eight Fe-peak element dispersions (6/8, \ion{Cr}{II}, \ion{Fe}{I}, Co, Ni, Cu and Zn) in NGC~288 and seven out of eight (7/8, \ion{Cr}{I}, \ion{Cr}{II}, \ion{Fe}{I}, Co, Ni, Cu and Zn) in NGC~362 (counting each ionisation state separately) cannot be explained by measurement errors and a spread in He. Of these, by our metric (1.5$\sigma_{\mathrm{ave+He}}$), four are considered marginally or truly genuine in NGC~288 (\ion{Fe}{I}, Co, Cu and Zn) and six are considered truly genuine in NGC~362 (\ion{Cr}{II}, \ion{Fe}{I}, Co, Ni, Cu and Zn).

While several GCs have been found to be inhomogenous in light elements and possibly Fe when measurement errors are on the order $>0.05$~dex \citep{meszaros20}, we interpret the spreads we observe as evidence that potentially \textit{all} GCs are inhomogeneous at the 0.02~dex level - measurable only through differential analysis. Because we have not selected for first or second generation stars only in our GCs, this could mean that potentially generation-independent spreads in metallicity exist in GCs at the level of 0.02~dex or above. This highlights the power of high precision measurements to answer questions like those posed in \citet{sneden05}, who explored the apparent existence of a [Ni/Fe] dispersion floor in GCs (at $\sim0.06$~dex). In our two GCs the values of $\sigma(\Delta^{\mathrm{Ni}}-\Delta^{\mathrm{Fe}})$ are 0.006~dex and 0.022~dex in NGC~288 and NGC~362 respectively, breaking through this proposed floor.

\subsubsection{Heavy Elements}
\label{sec:heavyelemdisp}
The final groups of elements we explore are the heavy elements, produced by way of the slow ($s$) and rapid ($r$) neutron capture processes. These elements are seeded from light nuclei through the addition of neutrons on varying timescales (slow or rapid when compared with $\beta$-decay of radioactive nuclei) \citep{meyer94, kappeler11, thielemann11}. The heavy elements we investigate are Sr, Y, Zr, Ba, La, Ce, Nd and Eu. 

The three elements Sr, Y and Zr are referred to as the first $s$-process peak and are produced in both AGB stars and through the weak $r$-process in electron-capture SNe and massive stars. The heavier elements, Ba, La and Ce belong to the second $s$-process peak and are predominately produced in AGB stars \citep{busso99, karakas14}. The remaining elements Nd and Eu have contributions from both $s$- and $r$-process sites, with Nd marginally dominated by the $s$-process and Eu considered an almost ``pure'' $r$-process element \citep{bisterzo11}. 
Sites of $r$-process formation include CCSNe, neutron star mergers (NSMs), neutron star-black hole mergers and magneto-rotational supernovae (MRSNe) \citep{cote18, chiaki20}. MRSNe are CCSNe that are triggered by strong rotation and/or magnetic fields \citep{cameron01, cameron03}. Ba and Eu are considered the characteristic $s$ and $r$-process elements respectively. 

\begin{figure}
	\includegraphics[width=\linewidth]{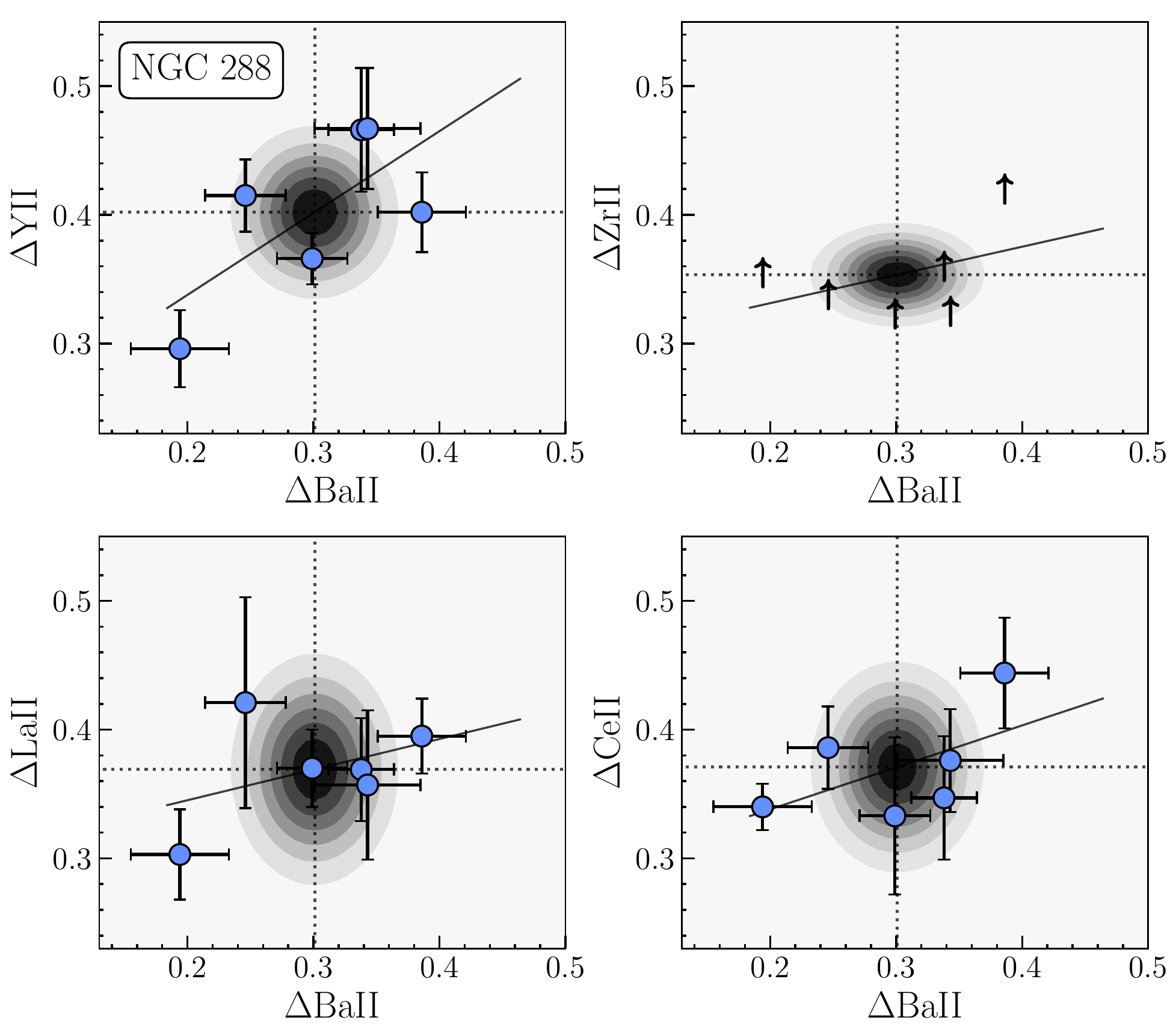}
	\includegraphics[width=\linewidth]{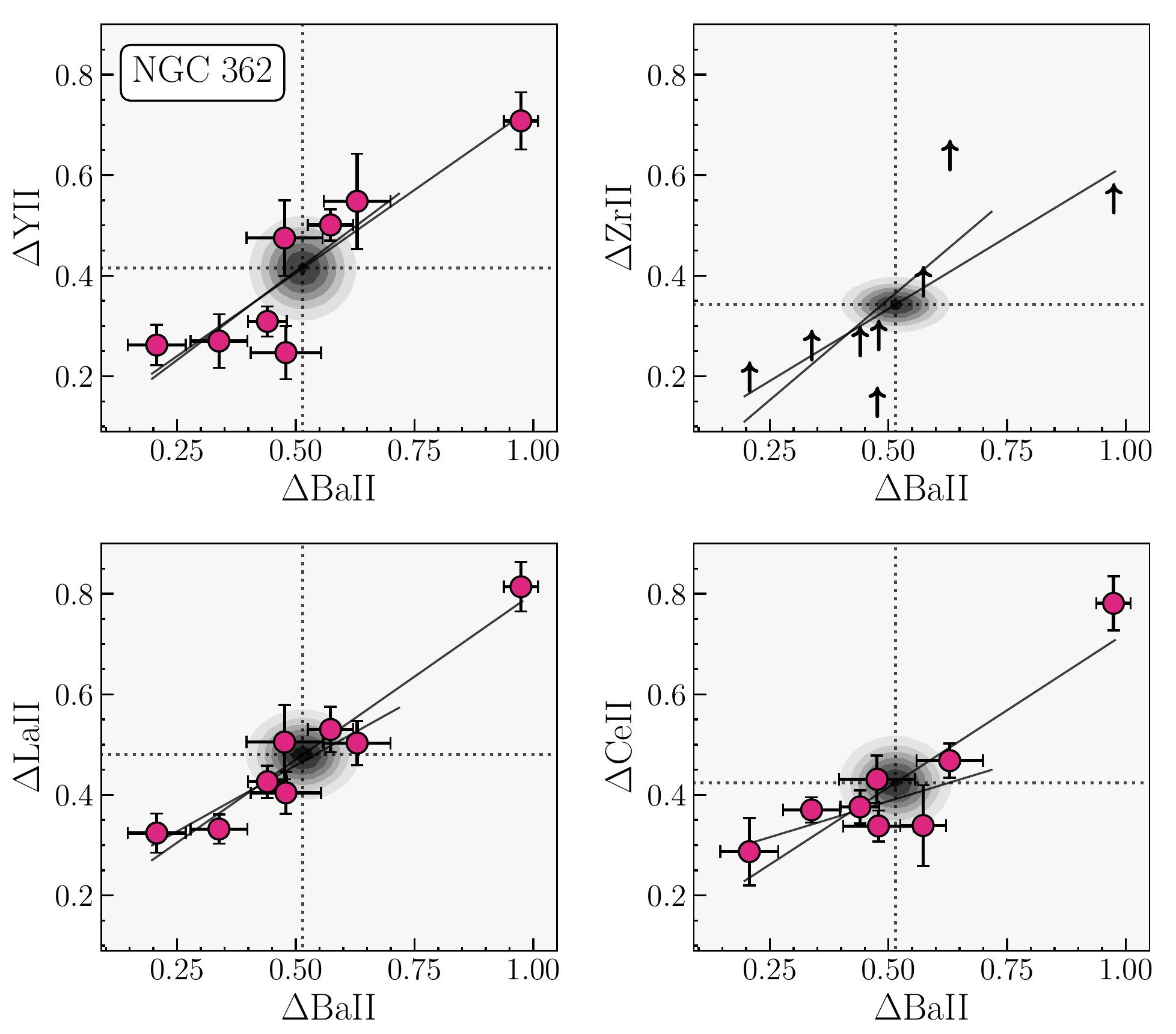}
    \caption{Trends in the $s$-process dominated elements $\Delta^{\mathrm{Y}}$, $\Delta^{\mathrm{Zr}}$, $\Delta^{\mathrm{Ce}}$ and $\Delta^{\mathrm{La}}$ as a function of $\Delta^{\mathrm{Ba}}$, the characteristic $s$-process element. Note that all four elements show positive correlation with Ba, supporting the interpretation of an $s$-process spread in both clusters. Contours representing the expected spread in a homogeneous distribution due to measurement errors alone are shown in each panel to highlight that the dispersion is genuine. Single line measurements are shown with an arrow.}
    \label{fig:bothsprocesscorr}
\end{figure} 

Fig.~\ref{fig:heavyelemdisp} reveals that both clusters show a spread in almost every $s$-process element measured. The spread is most significant in the well-measured elements Y ($\sigma_{\Delta^{\mathrm{Y}}}=0.06\pm0.02$ in NGC~288, $\sigma_{\Delta^{\mathrm{Y}}}=0.12\pm0.04$ in NGC~362)  and Ba\footnote{recall that the Ba lines are strong (EW$\sim150$~m\AA\,) but have all been hand-measured.} ($\sigma_{\Delta^{\mathrm{Ba}}}=0.06\pm0.02$ in NGC~288 and $\sigma_{\Delta^{\mathrm{Ba}}}=0.13\pm0.04$ in NGC~362). In the case of NGC~288, the spread in Y is $1.7\sigma_{\mathrm{ave+He}}$ and $1.8\sigma_{\mathrm{ave+He}}$ in Ba. In NGC~362, the spread is $2.1\sigma_{\mathrm{ave+He}}$ in Y and $2.0\sigma_{\mathrm{ave+He}}$ in Ba. Additionally, with the exception of Nd, there appears to be a trend of increasing dispersion with increasing $s$-process contribution in the elements shown in Fig.~\ref{fig:heavyelemdisp}. The $s$-process elements showing the largest dispersions, Sr, Y, Zr and Ba are dominated ($>80\%$) by the $s$-process \citep{bisterzo11}. While La and Ce with $s$-process contributions of $<80\%$ show decreased dispersion \citep{bisterzo11}.

Overall, NGC~362 shows more significant spreads in all the $s$-process elements. This is in agreement with what was found in the study of \citet{shetrone00} for Ba, the only $s$-process element measured in their study. Considering only stars in-common between the two studies, $\sigma_{\mathrm{Ba}}=0.08\pm0.02$ in NGC~288 and $\sigma_{\mathrm{Ba}}=0.16\pm0.05$ in NGC~362, neglecting NGC362-1441. Although the scale of the abundance errors is not noted in their study, the average $1\sigma$ errors appear to be $\sim0.10$~dex in their Fig.~4. Therefore, the $s$-process spread in NGC~362 may have been tentatively identified in the original study of \citet{shetrone00}. As NGC~362 is a Type II GC and the cluster has since been confirmed to host a spread in $s$-process elements \citep{carretta13, marino19}, we simply confirm this result. 

The average spread in $s$-process elements ($\Delta^{s-\mathrm{proc}}-\Delta^{\mathrm{Fe}}$, considering only Y, Ba, La, Ce and Nd) is $\sigma_{\Delta^{s-\mathrm{proc/Fe}}}=0.04\pm0.005$ for NGC~288, only slightly larger than the average error associated with measurement and a spread in He. In the case of NGC~362 the spread increases to $\sigma_{\Delta^{s-\mathrm{proc/Fe}}}=0.09\pm0.03$, which is $1.6\sigma_{\mathrm{ave+He}}$. If we consider all the $s$-process elements, the certainty in the dispersion being real remains unchanged for NGC~288 but increases to $2.3\sigma_{\mathrm{ave+He}}$ in NGC~362.

\begin{figure}
	\includegraphics[width=\linewidth]{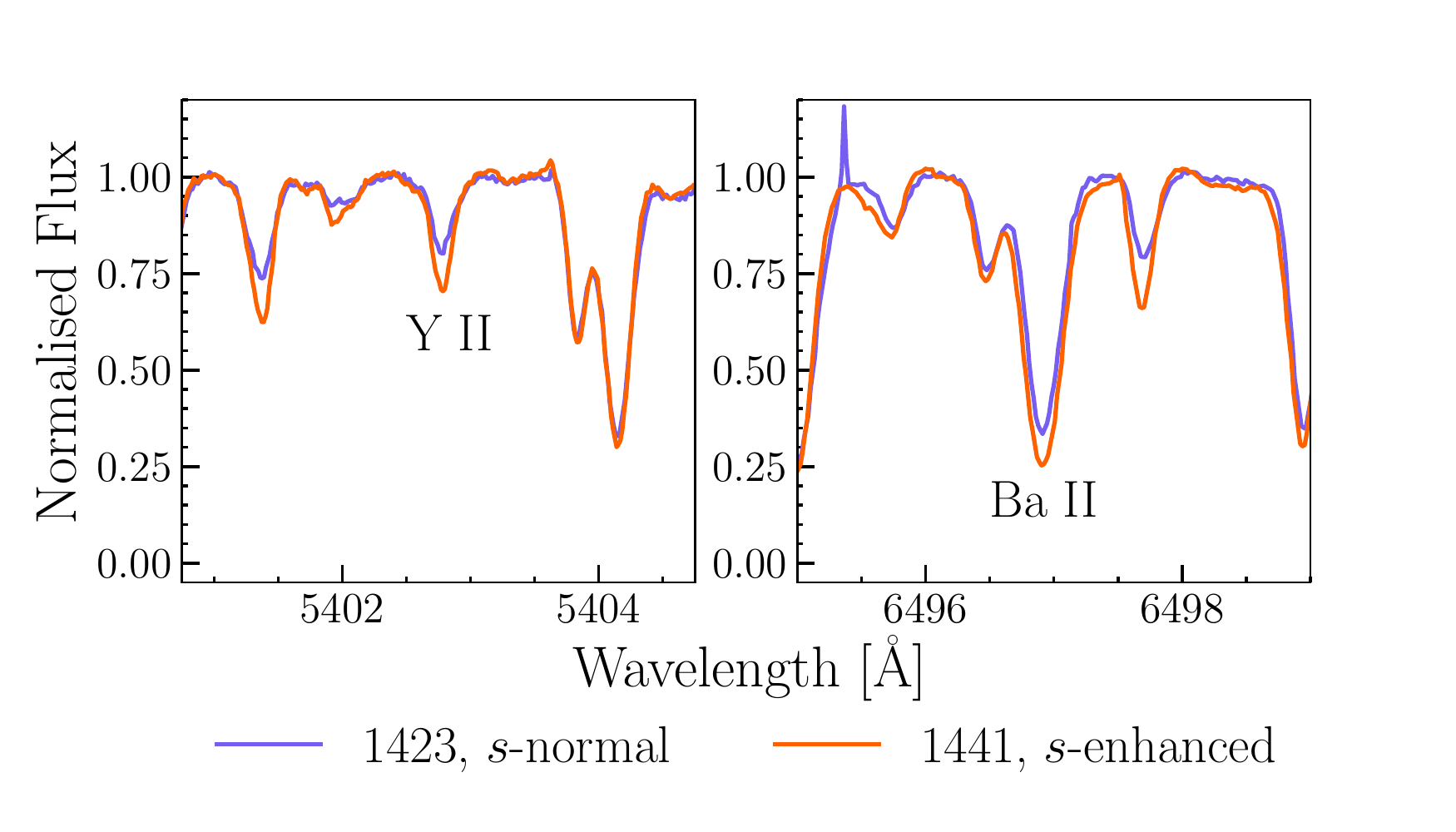}
    \caption{\todo{Spectral samples of the $s$-process normal star NGC362-1423 and the $s$-process rich star NGC~362-1441 centred around the 5402.77~\AA\, \ion{Y}{II} (\textit{left}) and 6496.90~\AA\, \ion{Ba}{II} (\textit{right}) lines. This comparison highlights the scale of $s$-process enhancement in NGC362-1441.}}
    \label{fig:spectralcomp}
\end{figure}

Although we are only able to measure a single \ion{Eu}{II} line at 6645.1\AA\,, we find a significant spread in the Eu abundance in NGC~362 ($\sigma_{\Delta^{\mathrm{Eu}}}=0.09\pm0.04$ or $2.4\sigma_{\mathrm{ave+He}}$) but not in NGC~288. Given the abundance errors on the order $\sim0.10$~dex, a spread in Eu in NGC~362 was not identified by \citet{shetrone00}. An $r$-process spread could explain the increase in dispersion in Nd ($\sim50\%$ $s$-contribution), at odds with the trend mentioned earlier for the $s$-process elements in NGC~362. In the case of the $r$-process spread ($\Delta^{r-\mathrm{proc}}-\Delta^{\mathrm{Fe}}$, considering only Eu), the spread in NGC~362 is $\sim1.5$ times larger than the measurement error plus He-spread. In the case of NGC~288, we do not consider the spread to be genuine by our metric, and given that it is on the order 0.02~dex it could be explained through the He scenario. 

To further validate the observed spread in $s$-process elements found in both clusters, we explore the correlations between the $s$-process dominated elements Y, Zr, La and Ce with Ba in Fig.~\ref{fig:bothsprocesscorr}. The correlations for NGC~288 are shown above the results for NGC~362. In both clusters we include contours associated with a homogeneous distribution in each element. That is, if all the stellar measurements were contained within the contours, then \textit{no} spread in that element would be detectable when considering measurement errors alone. The correlations also show positive trends between each element and the characteristic $s$-process element Ba. The discovery of a small but potentially significant spread in $s$-process elements in NGC~288 is unexpected given that it is a Type I GC. This could suggest that perhaps measurement uncertainties are hindering the detection of $s$-process spreads in other Type I GCs. 

Interestingly, we recover a single $s$-process-enhanced star in NGC~362, star 1441. The degree of enhancement is highlighted in Fig.~\ref{fig:spectralcomp} through the difference in line depths between 1441 (shown in orange) and the $s$-normal star 1423 (shown in purple). In Fig.~\ref{fig:bothsprocesscorr} we show the correlations with and without 1441, confirming that even without the presence of the $s$-process enhanced star, the correlations and spreads remain. Note that this is supported throughout this section as star 1441 has been neglected in the determination of all the elemental dispersions. Ba-rich, or generally $s$-process-enhanced stars are not uncommon in GCs, especially Type II GCs like NGC~362 which owe part of their classification to a spread in $s$-process element \citep{carretta13, milone17, marino19}. As discussed in Sec.~\ref{sec:intro}, NGC~362 is already known to harbour a $s$-process rich RGB, which \citet{carretta13} estimate contains $\sim6\%$ of the cluster stars. Given that we have selected eight stars in NGC~362, our discovery rate of 1/8 is roughly in line with this estimate. 

Using the HST ``magic'' filters, \citet{milone17} also estimated the fraction of first and second generation stars in Type II GCs based on the prevalence of the red vs. normal RGB. In NGC~362, \citet{milone17} estimated that stars from the red-RGB contributed 7.5$\pm$0.9\% to their sample of analysed stars. This is in closer agreement to our findings. In their discussion of red RGBs in Type II GCs, \citet{milone17} also cross-match their photometry of the Type II GC NGC~5286 with the high resolution spectroscopic catalogue of \citet{marino15}. They find that like in NGC~362, the red RGB primarily contains $s$-process (Ba) enhanced stars that are more metal-rich than the normal RGB stars. This is also the case for our $s$-process enhanced star, 1441, which is the most metal-rich star in our sample (see Tab.\ref{tab:stellarparams} for example). This suggests that 1441 is likely a member of the red RGB of NGC~362.

\begin{figure}
	\includegraphics[width=\linewidth]{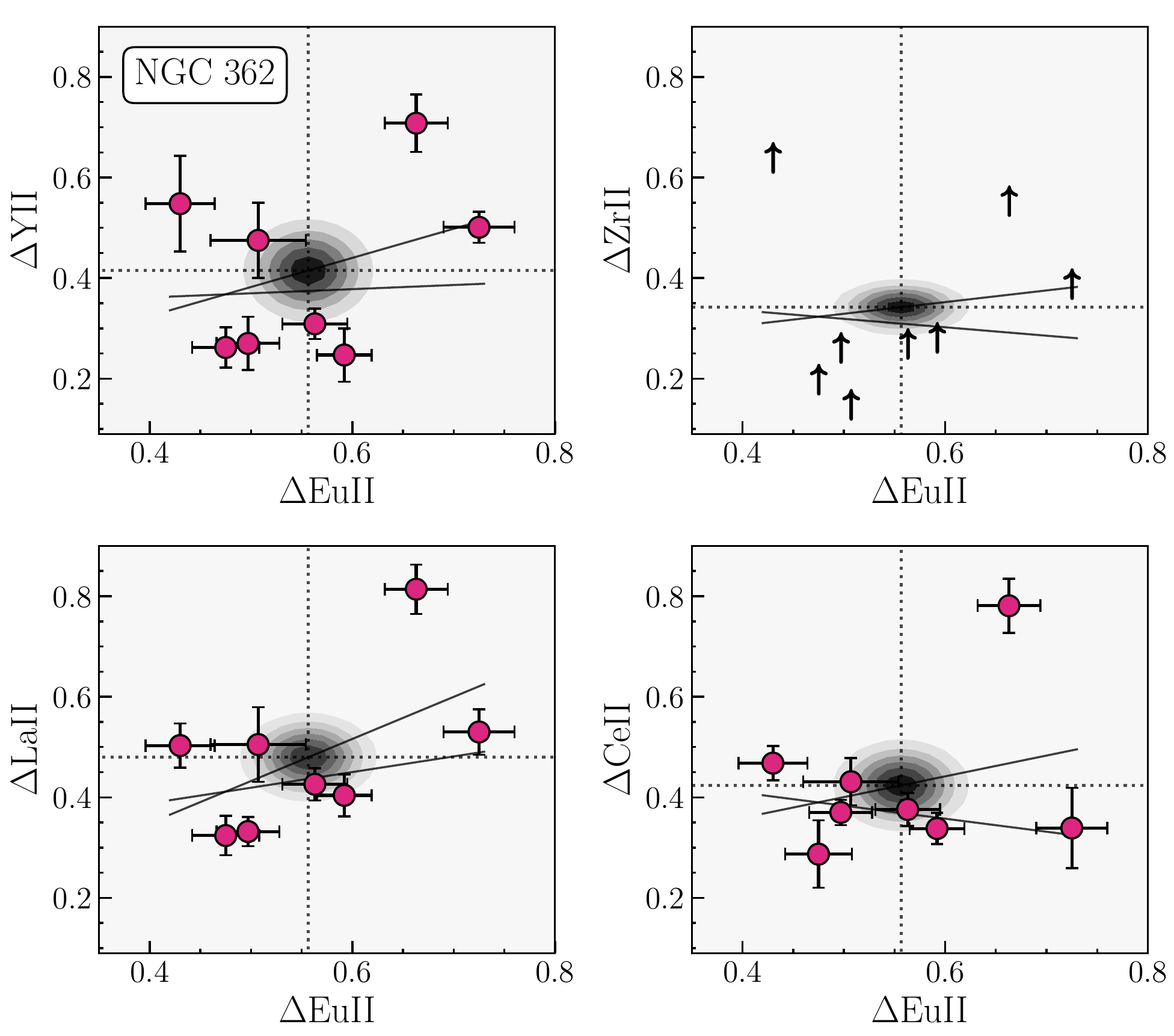}
    \caption{Trends in the $s$-process dominated elements $\Delta^{\mathrm{Y}}$, $\Delta^{\mathrm{Zr}}$, $\Delta^{\mathrm{Ce}}$ and $\Delta^{\mathrm{La}}$ as a function of the $r$-process element $\Delta^{\mathrm{Eu}}$. Note the decrease in strength of the correlations,  when compared to Fig.~\ref{fig:bothsprocesscorr} due to the presence of two distinct $s$-process groups in the cluster.}
    \label{fig:ngc362rscorr}
\end{figure}

To explore the $r$-process spread further in NGC~362, we re-plotted Fig.~\ref{fig:bothsprocesscorr} against Eu. As shown in Fig.~\ref{fig:ngc362rscorr}, no trends were present between any of the $s$-process elements and Eu ($r$-process) due to a clear split in the $s$-process abundances in our stars. The split was most visible in Y and La and is discussed in the next section. Referring to the work of \citet{roederer11}, specifically their Fig.~1 where they investigated $r$-process dispersion in GC literature data, the distribution of La vs Eu in NGC~288 most closely resembles clusters with ``small dispersion'' while NGC~362 resembles clusters with an ``$s$-process contribution''. In the case of NGC~362, other clusters occupying the same regime include two other Type II GCs, M~22 and $\omega$-Cen. 

Cross-matching \citet{roederer11} with the list of Type II GCs \citep[e.g. found in ][]{milone20}, the presence of an $r$-process spread is not as common among Type II GCs as an $s$-process spread. Whether some Type I GCs possess an $r$-process spread is unclear \citep[e.g. in case of M~92 ][]{roederer2011b, cohen11}. An $r$-process spread in GCs is interesting in the context of Galactic Archaeology \citep{freeman02} as \citet{roederer11} conclude that the existence of a spread could be primordial and is likely not the result of GC evolution. Thus a spread could potentially act as a unique tag to the GC formation environment, and act as a discriminator between accreted and in-situ Type I and II GCs. We revisit this in an upcoming companion paper. 

In total, we consider five out of eight (5/8, Sr, Y, Zr, Ba and Nd) heavy element spreads in NGC~288 and eight out of eight (8/8, Sr, Y, Zr, Ba, La, Ce, Nd, Eu) in NGC~362 to be genuine ($>1.5\sigma_{\mathrm{ave+He}}$). Considering both the spread from the best-measured $s$-process elements only and the spread in all $s$-process elements relative to H ($\sigma_{\Delta^{s-\mathrm{proc}}}$), both GCs show a global spread in $s$-process elements that cannot be explained by measurement errors and/or a spread in He.

\subsubsection{Two (or more?) $s$-process groups in NGC~362}
\label{sec:twosprocgroups}
From the distribution of Y, Ba and La in NGC~362, we find evidence for at least two distinct $s$-process groups within the cluster. The first group of $s$-process rich stars contains 1334, MB2 and 1401 and the extremely $s$-process enhanced star, 1441. The second group of $s$-process weak stars contains stars 1137, 2127, 1423 and 77. 

The split between the groups is most apparent in Y, as seen in the upper left panel of Fig.~\ref{fig:bothsprocesscorr}. This behaviour has also been observed in the GCs: M~2 \citep{yong14}, NGC~5286 \citep{marino15} and the massive disc cluster M~22, where a difference of 0.39~dex was found between the two $s$-process groups in the ratio of [La/Eu] \citep{marino11c}. This difference was recently confirmed using differential abundance analysis, providing further support for the result \citep{mckenzie22}. 
For our $s$-process rich group, using our best-measured $s$-process elements, we find an average $\Delta^{\mathrm{Y, Ba, La}}$ abundance of $0.62\pm0.15$ ($0.53\pm0.05$ without the $s$-process enhanced star, 1441), for the $s$-process weak group we find $\Delta^{\mathrm{Y, Ba, La}}=0.34\pm0.08$. This is a difference of $0.28\pm0.17$ ($0.19\pm0.09$ without 1441) between the two groups that cannot be explained by a spread in He. Cast as the ratio of $\Delta^{\mathrm{La}}-\Delta^{\mathrm{Eu}}$ (equivalent to [La/Eu]), we find average values of $0.007\pm0.13$ ($-0.05\pm0.11$ without 1441) in the $s$-process rich group and $-0.16\pm0.02$ in the $s$-process weak group. This is a difference between groups of $\sim0.17\pm0.13$ in $\Delta^{\mathrm{La}}-\Delta^{\mathrm{Eu}}$. Although the scale of the difference and the uncertainties are comparable, we tentatively treat them as separate groups in forthcoming discussions.

Two other interesting things to note about the two groups is that each contains an internal spread in Fe and that no universal offset in Fe is found between the two. Metallicity spreads in the first generation of GC stars is seemingly more ubiquitous, having been observed photometrically \citep{milone15, lardo22, legnardi22}, spectroscopically \citep{marino3201} and through simulations of GC formation \citep{mckenzie21}. The spread is largest in the $s$-process rich group, $\sigma_{\Delta^{\mathrm{FeI}}}=0.05\pm0.02$ ($\sigma_{\mathrm{meas+He}}=0.03$) compared to $\sigma_{\mathrm{Fe}}=0.01\pm0.004$ ($\sigma_{\mathrm{meas+He}}=0.02$) in the $s$-weak group. While the Fe-spread in the $s$-process weak group may be explained by measurement errors and a spread in He, the Fe-spread in the $s$-process rich group cannot. We revisit this in Sec.\ref{sec:ngc362unexpcorr}.

\subsection{Elemental Correlations: Group by Group}
\label{sec:elemcor_grouped}
To further explore the abundances of elements with common nucleosynthetic origins, we look for elemental correlations within each group discussed in the previous section. To measure the correlations we perform a linear fit to each element combination (i.e. $\Delta^{\mathrm{Y}}$ vs. $\Delta^{\mathrm{X}}$) and recover the uncertainty. The significance of the correlation is then cast as the ratio of the slope to the uncertainty. No instances of zero slope were found to disrupt this classification. The correlations for each group are presented in Appendix~\ref{app:allchemcorr} in Figs.~\ref{fig:alphapyramid}, \ref{fig:fepeakpyramid288}, \ref{fig:fepeakpyramid362}, \ref{fig:sprocesspyramid288} and \ref{fig:sprocesspyramid362}. For each element combination the slope of the correlation is also included in the associated tile.

Summarising the results of the group-by-group correlations, we find positive correlations for the majority of well-measured $\alpha$- and Fe-peak elements in both clusters. As a sanity check, the correlation between neutral Ti  (\ion{Ti}{I}) and singly ionised Ti (\ion{Ti}{II}) and neutral Fe  (\ion{Fe}{I}) and singly ionised Fe (\ion{Fe}{II}) are included and found to have strong positive correlations in both clusters. In general, trends in correlation and significance agree between both clusters in the $\alpha$- and Fe-peak elements. This is not the case for the heavy $s$- and $r$-process elements. NGC~288 shows little to no significant correlations between any of the heavy elements, while NGC~362 displays significant correlations between expected elements (e.g. Ba and La, quintessential $s$-process elements.) To investigate whether the higher number of significant correlations found in NGC~362 could be the result of the strongly $s$-process enhanced star 1441, we recreated Fig.~\ref{fig:sprocesspyramid362} without 1441 and found that the correlations remained, although they were slightly less significant. As with the other two element groups, no significant anti-correlations are seen between any of the $s$- and $r$-process elements.

\section{Unexpected Elemental Correlations: a Chemical Link to the Progenitor Environment?}
\label{sec:unexpeccorr}
We now explore the full range of elemental correlations and abundance ratios within each cluster. Unlike Sec.~\ref{sec:elemcor_grouped}, where we presented chemical correlations of elements with common nucleosythentic origins, we now look at correlations between elements from different groups and explore their respective ratios.

\subsection{Chemical Correlations}
\label{sec:chemcorr}
\begin{figure*}
	\includegraphics[scale=0.4]{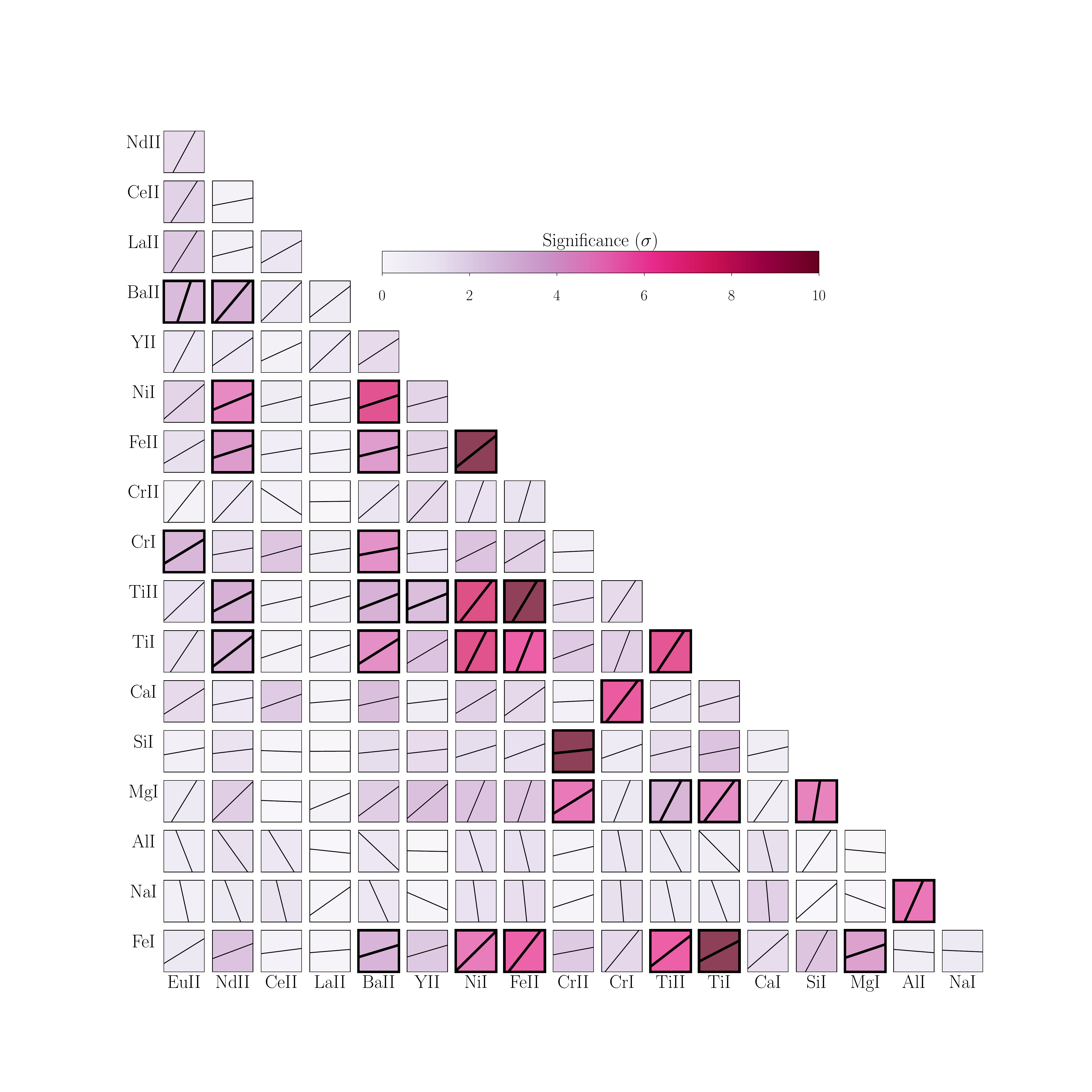}
	\caption{Elemental correlations coloured by statistical significance (with a maximum significance of 10$\sigma$) for a set of elements in this study in the GC \textbf{NGC~288}. The light elements with known star-to-star abundance variations (\ion{Na}{I}, \ion{Al}{I} and \ion{Mg}{I} have been excluded along with elements having only a single line measurement (with the exception of \ion{Eu}{II}). Elemental combinations with correlation strengths greater than $3\sigma$ are shown as the bolded boxes.}
    \label{fig:288pyramidplot}
\end{figure*}

\begin{figure*}
\includegraphics[scale=0.4]{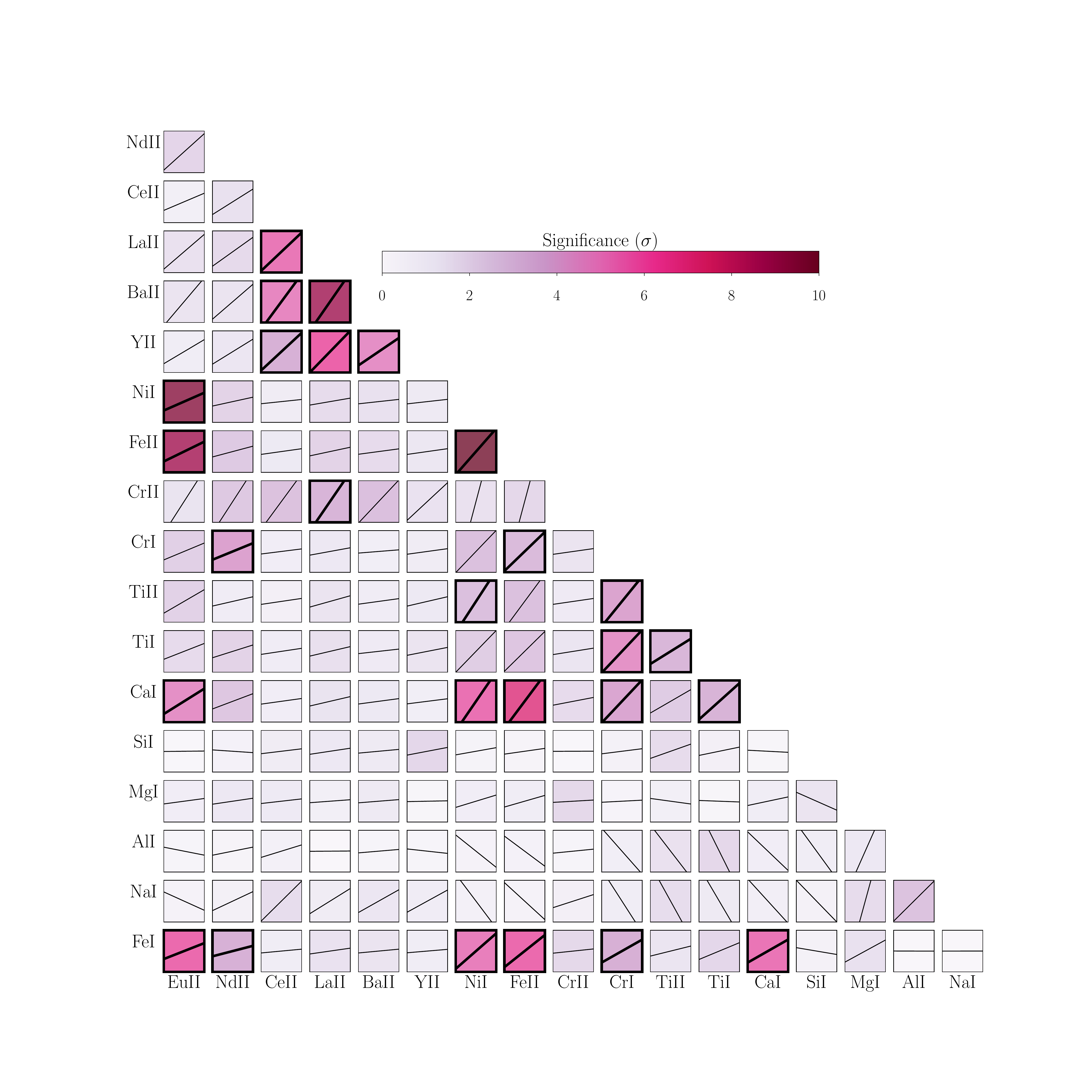}
    \caption{Same as Fig.~\ref{fig:288pyramidplot} for the GC  \textbf{NGC~362}, all stars are included except the $s$-process enhanced star NGC362-1441.}
    \label{fig:362pyramidplot}
\end{figure*}

All possible element combinations and correlations are shown in Fig.s \ref{fig:288pyramidplot} and \ref{fig:362pyramidplot} using the same conventions as those described in Sec.~\ref{sec:elemcor_grouped}. We exclude elements that have fewer than two individual stellar measurements, as these cannot be fit reliably. In total, each figure represents 105 unique correlations, presenting a daunting task of interpreting the significance of each. Thus, we choose to focus on elements with $3\sigma$ or greater correlation strength. This reduces the number of correlations in NGC~288 to 52 and to 54 in NGC~362. Note that in the case of NGC~362, the $s$-process enhanced star is included in the calculations of the correlations. We experimented with not including star 1441 and found no differences in what elements were correlated, only a decrease in the correlation strengths. The $3\sigma$-strength correlations are highlighted in Fig.s \ref{fig:288pyramidplot} and \ref{fig:362pyramidplot} as the bolded boxes. 

Two things are immediately apparent when considering the $3\sigma$ significant correlations. The first is that of the 105 unique combinations approximately 50\% of them exhibit correlations which are significant at the $3\sigma$ level or higher. This is further evidence suggesting that the clusters are inhomogeneous at the level at which we have measured them. The second thing that is apparent among the $3\sigma$-strong correlations is that all are positive in both clusters, \todo{this was also the case} in the study of \citet{yong2013}. 

Included in the inhomogenous chemical evolution scenario proposed by \citet{yong2013} to explain unexpected elemental correlations, is dilution of the intracluster medium with pristine gas. This would result in later generations of stars forming from gas enriched in every ratio of [X/H]. In this scenario, the light element variations in Na and Al are decoupled from the enrichment of elements heavier than Si \citep{yong2013}. Equal enrichment in every ratio of [X/H] equates to the prediction of an average slope of one for elements heavier than Si. Examining the correlations in Si and heavier, we find the average slopes of the correlations are $m=0.80\pm2.12$ and $m=0.86\pm0.84$ in NGC~288 and NGC~362 respectively. These averages increase to $m=1.26\pm1.06$ and $m=1.00\pm0.49$ for NGC~288 and NGC~362 respectively if only $3\sigma$ or stronger correlations are considered. Although the average values are close to one in both clusters, with or without a $3\sigma$-clip, the standard deviation in the set of slopes is too large to conclude whether enrichment by pristine gas is the source of the unexpected correlations in either cluster.

In the upcoming sections we discuss another scenario proposed by \citet{yong2013} included in the inhomogenous chemical evolution scenario, namely enrichment via a single nucleosynthetic source.

\subsubsection{NGC~288}
\label{sec:ngc288_corr}
At the $3\sigma$-level, there are several expected elemental correlations between elements from different groups (as we have classified them). An example of this is the correlation between the heavy $\alpha$-elements Ca and Ti (both \ion{Ti}{I} and \ion{Ti}{II}) and the Fe-peak elements Ni, Fe and Cr (in the case of Ca). This is because both Ca and Ti are also produced via SNIa \citep{chiaki20}. Additionally, the characteristic GC correlation between Na and Al is recovered. These are comforting results as they lend support to both our analysis and the unexpected correlations to be discussed. Note that we define ``unexpected'' correlations as correlations between elements that are \textit{not} produced via an obvious common nucleosynthetic channel.

We find two interesting trends among the significant unexpected correlations, the first being a general correlation between $s$-process elements and Fe-peak elements, and the second being a correlation between the $r$-process element Eu and the Fe-peak element Cr. In the case of the first trend, both the $s$-process elements Nd and Ba show trends with the Fe-peak elements Ni, Fe, Ti and Cr (in the case of Ba), while the $s$-process elements Ce, La and Y do not. Correlations between Fe-peak and $s$-process elements are a distinctive feature of many Type II GCs \citep[e.g. NGC~5286,][]{marino15}, yet another curiosity observed in our Type I GC.

The weighted average slope for the combinations of Fe-peak (Ni, Fe, Cr, Ti) vs. $s$-process (Ba, Nd) is $m=0.33\pm0.17$~dex. This implies that whatever process is enriching in $s$-process elements is more efficient than the Fe-peak production channel. Similarly, The slope of the correlation between Cr and Eu is $m=0.59\pm0.17$~dex. Finally, to assess the rate of $s$- to $r$-process production, we note that the best-fit to Ba vs. Eu was found to have a slope of $m=2.96\pm0.93$~dex.  

To consider inhomogenous chemical evolution in the proto-cluster environment as a potential explanation for the aforementioned trends, we must consider both, i) the possibility of a common nucleosynthetic site and ii) enrichment timescales. Beginning with a discussion of common nucleosynthetic sites, as discussed in \citet{chiaki20}, AGB stars dominate the production of Ba. They are also proposed to be responsible for the light element variations mentioned in Sec.~\ref{sec:lighelemintro}, through enriching the ISM between episodes of star formation \citep{dercole08, karakas10, dercole10}. However, the production of $s$-process elements is dominated primarily by long-lived, low-mass AGB stars in the mass range $\mathrm{M}\leq3-4$~M$_{\odot}$, while light element production is dominated by intermediate-mass AGB stars ($\mathrm{M}\geq3-4$~M$_{\odot}$) which reach temperatures high enough to facilitate hot bottom burning \citep[HBB, e.g. see][for a discussion of the different predictions for the two mass ranges.]{karakas09} Because of the differences in timescales between the two mass ranges, intermediate-mass AGB stars are thought to be more important in GCs \citep{karakasgcs}. We re-visit the difference in timescales in upcoming sections.

We find no correlation between Ba and any of the light elements we measured (Al, Na and Mg). Additionally, we find no significant correlations between Eu and Na, nor Ba and Al and instead, find anti-correlations between both sets of elements. Although, we do find a statistically insignificant correlation between Mg and Eu (potentially explained by the $r$-process contribution from CCSNe). The presence of a similar trend in 11 MW GCs, namely that no correlations existed between variations in light elements and La nor Eu, led \citet{roederer11} to their conclusion that some clusters harbour primordial spreads in $r$-process elements (as mentioned in Sec.~\ref{sec:heavyelemdisp}). This could suggest that the $s$- and $r$-process abundances we see are indicative of the proto-cluster environment. This then raises the question of whether a common nucleosynthetic site for both the Fe-peak elements and $s$- and $r$-process elements exists.

One potential site for the nucleosynthesis of both groups is through HNe, introduced in Sec.~\ref{sec:fepeak_disp}. HNe enrich in more $\alpha$ and Fe-peak elements than conventional CCSNe \citep{chiaki20}. MRSNe introduced in Sec.~\ref{sec:heavyelemdisp} as an $r$-process site are also a type of HNe and have been shown to reproduce the $r$-process abundance patterns in metal-poor stars \citep{yong21}. In their galactic chemical evolution model, \citet{chiaki20} find that only 60\% of Cr, Ni and Fe is produced via Type Ia enrichment, while the remaining 40\% is produced via HNe. 

HNe are also a proposed site for early $r$-process enrichment prior to the onset of NSMs. \citet{chiaki20} find that a minimum contribution of 3\% enrichment from HNe is needed to explain the abundance of Eu to reproduce observations of MW stars. \citet{chiaki20} also find that HNe are needed to reproduce the abundance of Ba at very early times, before the onset of AGB stars. The timescale for the enrichment of HNe matches with the picture of primordial enrichment of proto-cluster environment as it occurs at the earliest times. Therefore, HNe could serve as the nucleosynthetic link enriching the proto-cluster environment in both Fe-peak and $s$/$r$-process elements.

\subsubsection{NGC~362}
\label{sec:ngc362unexpcorr}
As discussed in Sec~\ref{sec:twosprocgroups}, we found two distinct $s$-process groups within NGC~362. Therefore, we consider the elemental correlations in NGC~362 using i) all the stars (shown in Fig.~\ref{fig:362pyramidplot}), ii) only using stars from the $s$-process rich group (shown as the top panel in Fig.~\ref{fig:362pyramidplot_sproc}) and iii) stars from the $s$-process weak group (the bottom panel in Fig.~\ref{fig:362pyramidplot_sproc}). Beginning with Fig.~\ref{fig:362pyramidplot}, as in the case of case of NGC~288 correlations which are significant at the 3$\sigma$ level are found among many expected elements. These include correlations between the $s$-process elements Y, Ba and Ce and correlations between the Fe-peak elements Ca (some SNIa contribution), Ti, Ni and Fe. Again, as seen in NGC~288, a correlation is found between the Fe-peak elements Ni and Fe and the $r$-process element Eu and anti-correlations are found between Eu and the light elements. Contrary to what was seen in NGC~288, no strong correlations are found between the Fe-peak elements and the $s$-process elements, except in the case of Cr (both ioinisation states) and La and Nd. 

Splitting the cluster into the two $s$-process groups reveals some interesting differences. Looking first at the top panel of Fig.~\ref{fig:362pyramidplot_sproc}, showing the $s$-process rich group ($s$-rich), two notable differences appear compared to Fig.~\ref{fig:362pyramidplot}. The first being further correlations between Fe-peak elements (Ti and Cr) and Eu, and the second being new correlations between the light element Al and $s$-process elements Ce, Ba and Y. While in the bottom panel of Fig.~\ref{fig:362pyramidplot_sproc}, showing the $s$-process weak ($s$-weak) group we find correlations between $s$ and $r$-process elements in addition to anti-correlations between the light elements Na and Al and the $\alpha$-elements Ca and Ti. The other major difference between the two groups is the presence of a strong Mg-Si anti-correlation in the $s$-process rich group and not in the $s$-process weak group. Both groups show the characteristic Na-Al correlation and spreads in the light elements. However, the $s$-rich group shows a significantly larger spread in the light elements; Na, Mg, Si and Ca compared to the $s$-poor group. The only exception to this is Al, where the $s$-weak group displays a larger dispersion.

Two things lead us to believe that the two $s$-process groups represent different stellar populations and thus potentially different epochs of star formation. The first is the existence of significant correlations between Al and the $s$-process elements Ce, Ba and Y in the $s$-rich group and \textit{not} the $s$-weak group. The second point is the reversal of the sign in the Mg-Si correlation between groups. The $s$-rich group shows a $3\sigma$ strong \textit{anti}-correlation between Mg and Si, while the $s$-weak group shows a weak \textit{correlation} between Mg and Si. The $s$-rich group is also more enhanced in Si compared to the $s$-weak group (by $\sim0.1$~dex).

As discussed in Sec.~\ref{sec:ngc288_corr}, intermediate mass AGB star enrichment of the ISM could explain the light element (anti-)correlations observed through HBB at different temperatures. As a result of a combination of the CNO cycle and NeNa and MgAl burning, gas from intermediate-mass AGB stars is expected to be rich in Na and Al and weak in O and Mg. Further down in the mass spectrum, low-mass AGB stars are candidate primary sources of $s$-process enrichment. If we assume some overlap in the production of both light elements \textit{and} $s$-process elements in the overlapping mass region between low- and intermediate-mass AGBs, this could explain the strong Al-$s$-process correlations we observed in the $s$-rich group. The Mg-Si anti-correlation in the $s$-rich group could also be the result of enrichment of the ISM via massive, evolved AGB stars. This occurs through leakage in the MgAl chain at high temperatures producing Si \citep{karakas03, ventura09, ventura12}. 

Under the assumption that the $s$-rich group formed from more AGB-enriched material, this implies that the $s$-rich group could be \textit{younger} than the $s$-weak group. However, two things complicate this picture, the first is the lack of Fe-evolution between the two groups (there is no obvious difference in metallicity between the two groups, with or without the $s$-process enhanced star 1441), and the second is the presence of a Na-Al correlation in \textit{both} groups. A lack of Fe-evolution implies no obvious contribution from the most rapid enrichment pathway, namely CCSNe. While the presence of a Na-Al correlation in both groups implies that intermediate-mass AGB stars have enriched both populations.

The first complication is likely a result of our selection of stellar siblings and may not be present in a study with a larger sample size. In the study of the massive Type II GC M~22 by \citet{marino11c} (introduced in Sec.~\ref{sec:twosprocgroups}) which included chemistry of 35 RGB stars, they saw a metallicity difference between their two $s$-process groups. On average the $s$-rich group was more metal-rich than the $s$-weak group. However, they also saw an overlap between the two groups of $\sim0.1$~dex in [Fe/H], the same range spanned by our sample. 

Regarding the second complication, the presence of Na-Al correlations in both groups (or equivalently Na-O anti-correlations), \citet{marino11c} also saw this in both $s$-process groups in M~22. To explain both the timescales and enrichment patterns seen in both populations, \citet{marino11c} presented a formation theory. The theory suggests a formation scenario in which initial enrichment via high velocity CCSNe ejecta occurred to increase the overall abundance of Fe in the outer parts of the cluster. This was then followed by the injection of light elements via intermediate-mass AGB stars. The $s$-weak population is then proposed to have formed in the inner region of the cluster from the intermediate-mass AGB ejecta. After the formation of the $s$-weak population, pollution of the ISM via low-mass AGB occurred introducing $s$-process elements. Ultimately, the $s$-rich population is proposed to have formed (also in the inner region of the GC) from both low-mass AGB-enriched material and CCSNe ejecta that fell into the GC centre following cooling. Note that this scenario also requires some production of the light elements by low-mass AGB stars to introduce the light element (anti-)correlations seen in the $s$-rich stars.

The scenario presented in \citet{marino11c} could also explain what we observe in NGC~362, with the exception of enrichment via CCSNe to explain metallicity variations. This assumes that no metallicity variations exist between the populations based off our small sample (as discussed this may not be the case.) However, this assumption is supported by the \textit{lack} of any trend in the abundance of Ba as a function of metallicity in NGC~362 seen in the studies of \citet{shetrone00} and \citet{carretta13}. \citet{carretta13} did report statistically significant, but weak, anti-correlations between Ba and the light elements Mg and O, also in agreement with our findings. Recall that we also recover a $3\sigma$ significant anti-correlation between Ba and Mg in our $s$-rich group. \citet{carretta13} did not report any complementary correlations between Ba and Si, nor between Ba and Na. 

Massive binaries have also been proposed to explain the production of light element (anti-)correlations \citep{demink09, renzini22}. In their model, \citet{renzini22} find massive binaries can enrich the intracluster medium with CNO products/light elements during the common envelope phase, and prior to the stars detonating via CCSNe. \citet{renzini22} also cite the relatively gentle expulsion of the stellar envelopes as support for the theory, as it allows the GC to retain the processed material. One of the major benefits of this theory is the rapid production of light element (anti-)correlations within the cluster. Massive binaries are predicted to enrich on $\sim10$~Myr timescales vs $\sim100$~Myr timescales as predicted from intermediate-mass AGB models \citep{fishlock14, shingles14}. However, enrichment in heavy $s$-process elements like Ba cannot be explained by massive binaries \citep{renzini22}. Perhaps early enrichment by massive binaries could explain the light element variations we see in our $s$-weak stars, while contributions from intermediate-mass AGB stars could explain the variations seen in the (younger?) $s$-rich population. Low-mass AGB stars remain the $s$-process polluter between generations in our scenario.

In the case of NGC~362 \citep[as in the case of most Type II GCs, eg. in NGC~1851, NGC~5286 and M~22][]{marino14, marino15, marino11c}, the picture of formation of the different chemical populations is incredibly complex. No single population of polluters (e.g. massive binaries, CCSNe, or AGB stars), can explain the chemical differences we observe between the two populations. Constraining any age differences between the two populations (if they exist) would greatly clarify the picture.

Bearing the previous discussion in mind, if our classifications of the two generations are correct, the potential spread in Fe in the $s$-weak stars discussed in Sec~\ref{sec:twosprocgroups} implies a primordial spread in Fe within the first generation of stars in NGC~362 \citep[as observed in simulations of GC formation, ][]{mckenzie21}. After removing the (anti-)correlations explained through normal cluster evolution and separate episodes of star formation, the correlation between the Fe-peak elements and Eu remains in the $s$-rich population and is of greater significance than could be explained by the [X/H] contribution from He of 0.015~dex. Additionally, the Eu abundance remains the same between groups - implying it is not the result of cluster evolution \citep[this is also the case in the two $s$-process groups in M~22][]{marino11c, mckenzie22}. As in NGC~288, this could imply a common nucleosynthetic origin for the two, one possible site being HNe. Again, we propose that the $r$-process abundance spread is primordial, as a spread in Eu is found in both $s$-groups.

\subsubsection{Comparing NGC~288 and the $s$-process weak population in NGC~362}
Given that the $s$-weak group in NGC~362 was found to share many similarities with NGC~288, we now compare the two. The most significant shared correlation is found between Fe-peak elements and $s$-process elements. This trend is more significant in NGC~288, but can be found between Cr, Fe and La and Cr and Ba in the $s$-weak group within NGC~362. As discussed in Sec.~\ref{sec:ngc288_corr}, this could be due to a shared production site, namely HNe. Given that the clusters share some common and unexpected correlations, it could be that they formed out of the same pristine gas around the same progenitor, enriched by the presence of HNe. Yet, they do not exhibit identical correlations, indicating that perhaps they formed from discrete pockets of gas located within the same galaxy. In our upcoming companion paper we perform a comprehensive comparison of the two clusters and find both striking similarities and differences in various elements.   

\section{Summary \& Conclusions}
\label{sec:sumconc}
From the original study of \citet{shetrone00}, we have re-analysed six stars from NGC~288 and eight stars from NGC~362 using high quality UVES spectra. The stars were selected to act as ``stellar siblings'', sharing similar values of effective temperature (T$_{\mathrm{eff}}$), surface gravity (log~$g$) and metallicity ([Fe/H]). Differential abundance analysis was performed on all the stars using the same reference star from the study of \citet{yong2013}. This technique was chosen to minimise uncertainties in the final (differential) abundances, leading to the successful recovery of errors on the order 0.01-0.02 dex in several well-measured elements. Ultimately, the abundance precision was only limited by the S/N of the data (i.e., accuracy of EWs) and the range of stellar parameters in the program stars relative to the reference star.

From these abundances, statistically significant intra-cluster spreads were found in both clusters in several light elements, Fe, Ni and several heavy elements - many of which could not be explained by an ab-initio spread in He within the clusters. Significant correlations were also found between elements sharing common nucleosynthetic origins and between those that do not. We propose HNe as a common nucleosynthetic site to explain the correlations between Fe-peak and $s/r$-process elements found in both GCs. \todo{In the future, large aperture telescope coupled with efficient high resolution spectrographs could probe whether these dispersions and correlations extend down the RGB and potentially into the main sequence.}

Two $s$-process groups were found within NGC~362, most clearly separated in Y. The first, $s$-weak group, is comprised of stars 1337, 2127, 1423 and 77. The second, $s$-rich group is comprised of stars 1334, MB2, 1401 and the highly $s$-process enriched star 1441. An average offset of $0.34\pm0.08$~dex was found in $\Delta^{\mathrm{Y, Ba, La}}$ when including star 1441. Given the presence of $3\sigma$, or stronger, correlations between Al and the $s$-process elements Y, Ba and Ce and the presence of a MgSi anti-correlation in the $s$-rich group, we theorise that the $s$-rich group formed after the $s$-weak group. Enrichment of Al and $s$-process elements, coupled with the depletion of Mg via leakage in the MgAl chain to produce Si, lead us to propose AGB stars as the primary polluter between the two episodes of star formation - if an overlap between low- and intermediate-mass AGB stars exists \citep{karakas03, dercole08, dercole10}. The $s$-weak group is dominated by the $r$-process, with a $\Delta^{\mathrm{La}}-\Delta^{\mathrm{Eu}}$ (analogous to [La/Eu]) ratio of $-0.16\pm0.02$. Under the assumption that the $s$-weak group represents the an earlier epoch of star formation in the cluster, we propose that this $r$-process abundance is primordial. \smallskip

\noindent In the following bullet points we present the main findings of the study.

\begin{itemize}
    \item Utilising differential abundance analysis techniques we recovered average abundance errors on the order 0.01 - 0.02~dex in the clusters NGC~288 and NGC~362. The best measured element in both clusters was \ion{Fe}{I}, with average uncertainties of 0.010 and 0.016~dex in NGC~288 and NGC~362 respectively.
    \item Genuine dispersions in several elements were found in both clusters (12/22 in NGC~228 and 19/22 in NGC~362). We define a       ``genuine'' spread to be larger than 1.5 times the quadrature sum ($\sigma_{\mathrm{ave+He}}$) of the average measurement error and uncertainty introduced by an \textit{ab-initio} spread in He abundances (0.009~dex in NGC~288 and 0.015~dex in NGC~362).
    \item Of the light elements, a genuine spread was found in Na, Al and Ti in NGC~288 and in Na, Al, Si, Ca and Ti in NGC~362. A genuine spread in average $\alpha$-elements ($\sigma_{\Delta^{[\alpha\mathrm{/Fe}]}}$) of 0.05~dex is detected in NGC~362 and a tentative detection of a 0.02~dex spread is found in NGC~288, at odds with its definition as a Type I GC.
    \item Genuine spreads in Ni and \ion{Fe}{I} are detected in NGC~362 at the level of $\sigma_{\Delta^{\mathrm{Ni}}}=0.039\pm0.011$ and $\sigma_{\Delta^{\mathrm{Fe}}}=0.035\pm0.010$, or $1.5\sigma_{\mathrm{ave+He}}$ in both cases. Potentially significant spreads in Ni and \ion{Fe}{I} are also detected in NGC~288 at the level of $\sigma_{\Delta^{\mathrm{Ni, Fe}}}=0.021\pm0.007$ or $1.2\sigma_{\mathrm{ave+He}}$ in the case of Ni and $1.4\sigma_{\mathrm{ave+He}}$ in the case of \ion{Fe}{I}. To the best of our knowledge, this is the first time a significant spread in Fe has been found in NGC~288.
    \item Given the above point, we suggest that all GCs are inhomogeneous at the 0.02~dex level, highlighting the power of differential abundance techniques.
    \item Both clusters display genuine spreads in $s$-process elements, which cannot be explained by a spread in He (in Sr, Y, Zr, Ba and Nd in NGC~288 and Sr, Y, Zr, Ba, La, Ce, and Nd in NGC~362. 
    \item The largest abundance spread seen in NGC~288 is in Y at the level of $\sigma_{\Delta^{\mathrm{Y}}}=0.059\pm0.019$ or $1.7\sigma_{\mathrm{ave+He}}$. The largest spread in NGC~362 (not from a single line measurement and neglecting star 1441) is seen in Ba at the level of $\sigma_{\Delta^{\mathrm{Ba}}}=0.131\pm0.038$ or $2.0\sigma_{\mathrm{ave+He}}$. A spread in $r$-process elements is found in NGC~362 but not NGC~288. 
    \item We find at least two distinct $s$-process groups in NGC~362, separated by 0.3~dex in $\Delta^{\mathrm{Y, Ba, La}}$ and aided by the presence of an extremely $s$-process enhanced star. 
    \item Given both the presence of strong correlations between Al and several $s$-process elements, and a significant positive correlation between Mg-Si in the $s$-process rich group, we hypothesise that the $s$-rich group is younger than the $s$-weak group. This is in agreement with enrichment scenarios from AGB stars, if there is overlap between low- and intermediate-mass AGB star enrichment.
    \item In NGC~288 a $3\sigma$ or greater correlation is found between the $s$-process elements Ba, Nd and Fe-peak elements Ni and Fe. This trend is also observed in the $s$-weak population in NGC~362. 
     \item The $s$-weak population is dominated by the $r$-process, displaying a $\Delta^{\mathrm{La}}-\Delta^{\mathrm{Eu}}$ (analogous to [La/Eu]) ratio of $-0.16\pm0.06$. If the $s$-weak population truly represents an earlier epoch of star formation in NGC~362, this suggests a primordial origin for the $r$-process enrichment.
\end{itemize}

These results have significant implications for our understanding of globular cluster formation in the early Milky Way. In particular, our analysis provides evidence that NGC~288 and NGC~362 are chemically inhomogeneous in elements from all four nucleosythentic groups ($\alpha$-elements, Fe-peak, $s$- and $r$-process). By extension, we speculate that perhaps all GCs could be chemically inhomogeneous at the 0.02~dex level. Whether these inhomogeneities are primordial or the result of internal evolution remains to be seen. Regardless, these results act as a high-precision reproducible for theories of both Type I and II GC formation and place constraints on the level of (in)homogeneity in the ISM in the earliest dGal environments.

\section*{Acknowledgements}
\todo{We thank the referee, Chris Sneden immensely for his comments and suggestions that improved this paper.} A portion of this paper was written while SM had COVID-19. She is incredibly grateful to the amazing scientists who created a vaccine that kept her safe and healthy during that time. She also dedicates this paper to her best friend, a front-line nurse who despite everything continues to show up every day for her patients, and to her grandmother who lost her life to COVID-19 early in the pandemic. If she could have taken the vaccine she would have. She also wishes to acknowledge the traditional custodians of Mt. Stromlo, the Ngunnawal and Ngambri people and pay her respect to Elders past and present. SM acknowledges funding support from the the Natural Sciences and Engineering Research Council of Canada (NSERC), [funding reference number PGSD3 - 545852 - 2020]. Cette recherche a été financée par le Conseil de recherches en sciences naturelles et en génie du Canada (CRSNG), [numéro de référence PGSD3 - 545852 - 2020]. Funding for the Stellar Astrophysics Centre is provided by The Danish National Research
Foundation (Grant DNRF106).

This work relies heavily on the Astropy \citep{astropy1, astropy2}, SciPy \citep{scipy}, NumPy \citep{numpy} and Matplotlib \citep{matplotlib} libraries. Based on observations collected at the European Organisation for Astronomical Research in the Southern Hemisphere under ESO programme 075.D-0209(A).
\section*{Data Availability}
The data underlying this article are available in the article and online
through provided links and supplementary material.



\bibliographystyle{mnras}
\bibliography{accepted} 




\appendix

\section{Verifying Reference Star Consistency}
\begin{figure}
	\includegraphics[width=\linewidth]{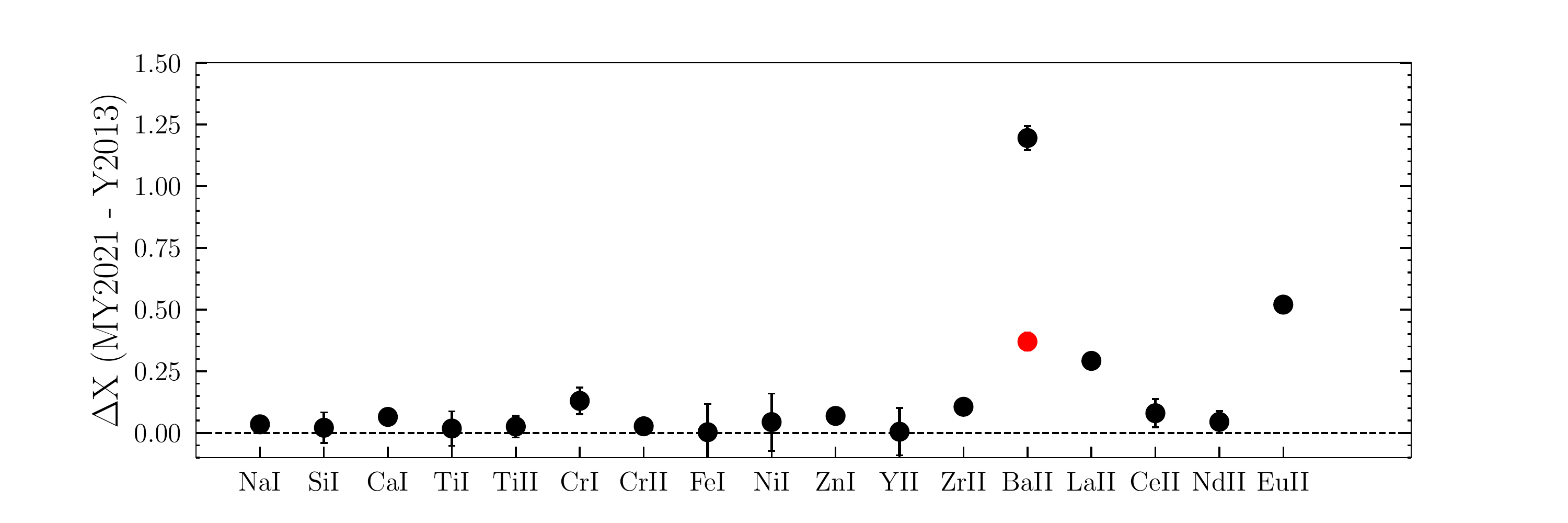}
    \caption{Absolute abundance differences measured in the reference star mg9 for the elements in common between this study and \citet{yong2013}. Almost all elements show agreement within $1\sigma$ with the exception of \ion{Ba}{II}, \ion{La}{II} and \ion{Eu}{II}. This is likely due to the use of spectral synthesis in the original study for the elements \ion{La}{II} and \ion{Eu}{II} . In the case of \ion{Ba}{II}, HFS corrections were applied in this study and not in Yong2013. The red marker shows the difference if HFS corrections are not applied. Note that the same stellar parameters were assumed for the reference star in both studies.}
    \label{fig:referencestardif}
\end{figure}

To perform our differential abundance analysis, we re-analysed the reference star mg9 from the study of \citet{yong2013} following the procedure described in Sec.~\ref{sec:obsanalysis}. A comparison of the recovered elemental abundances between studies (the difference between the two) is shown in Fig.~\ref{fig:referencestardif}. The large difference in \ion{Ba}{II} abundances between the two studies can be partially explained by the lack of HFS corrections applied to the element in \citet{yong2013}. To investigate this we re-determined the \ion{Ba}{II} abundance without applying HFS corrections and re-calculated the difference. Without HFS corrections the difference between studies is shown as the red point, significantly reducing the tension between the two.

\section{Additional Chemical Correlations}
\label{app:allchemcorr}
\begin{figure}
	\includegraphics[scale=0.4]{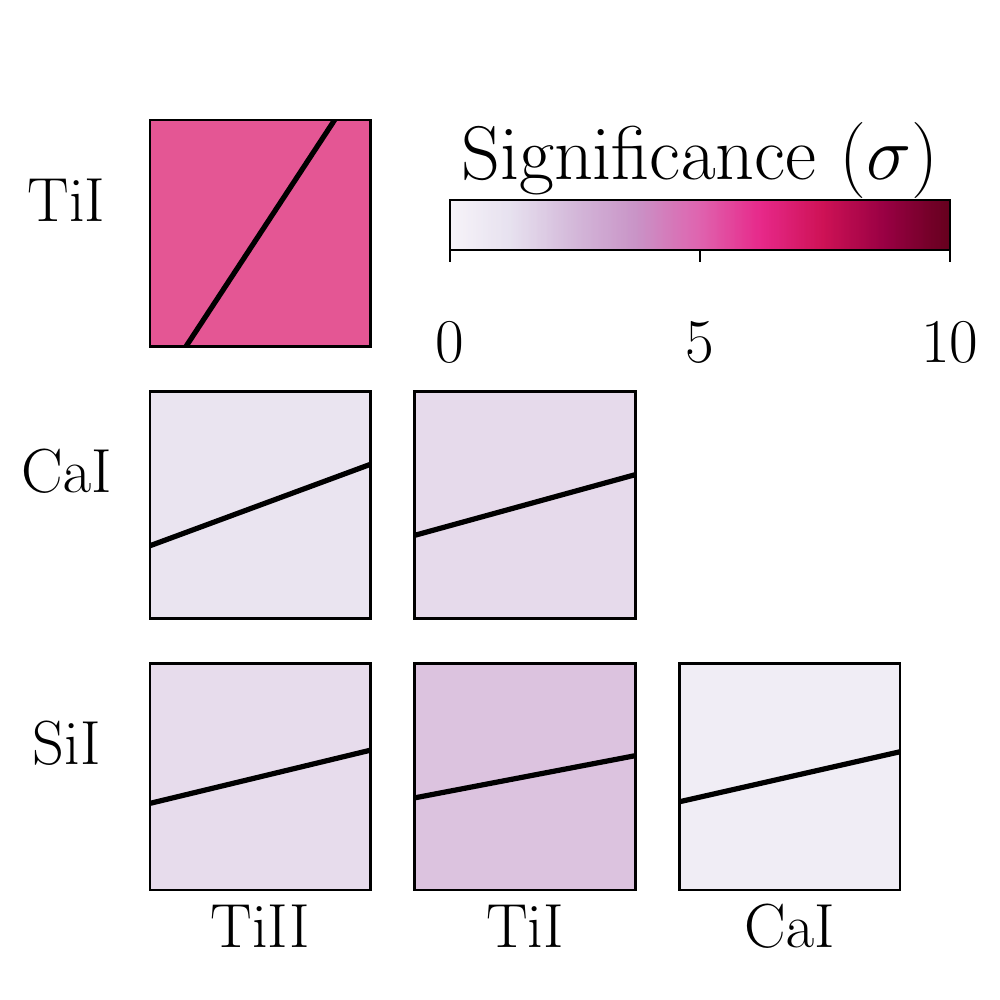}
	\includegraphics[scale=0.4]{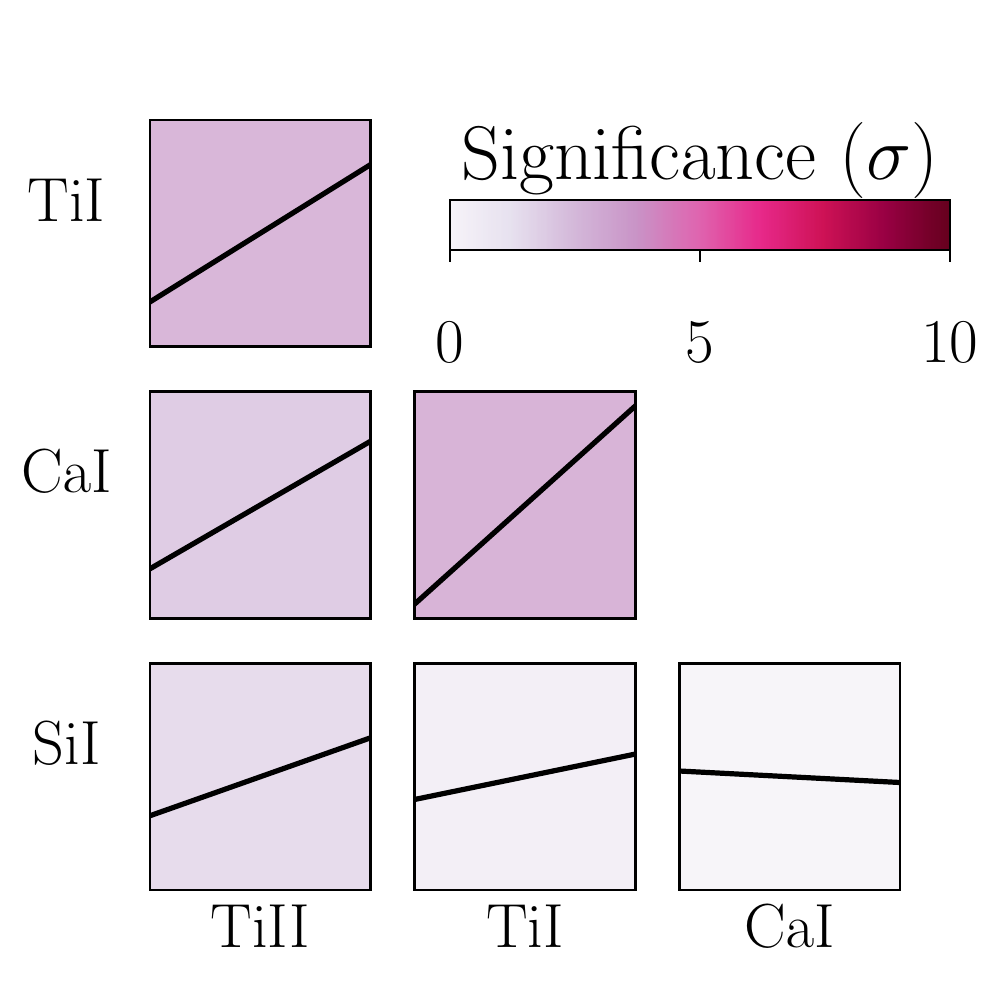}
	\caption{$\alpha$-element correlations coloured by statistical significance for \textbf{NGC~288 (left)} and \textbf{NGC~362 (right).}}
    \label{fig:alphapyramid}
\end{figure}

\begin{figure}
	\includegraphics[scale=0.4]{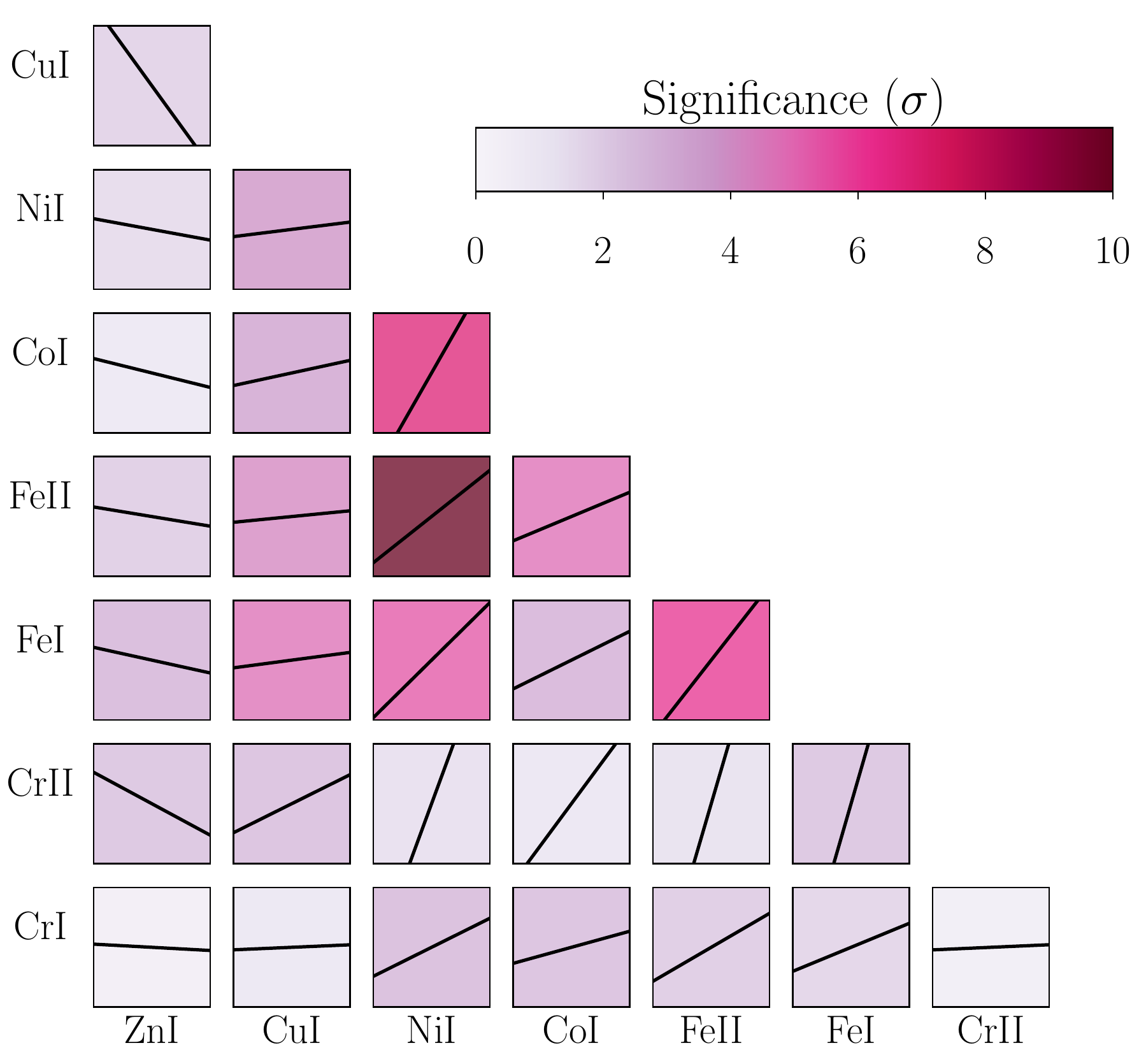}
	\caption{Fe-peak element correlations coloured by statistical significance for \textbf{NGC~288}.}
    \label{fig:fepeakpyramid288}
\end{figure}

\begin{figure}
	\includegraphics[scale=0.4]{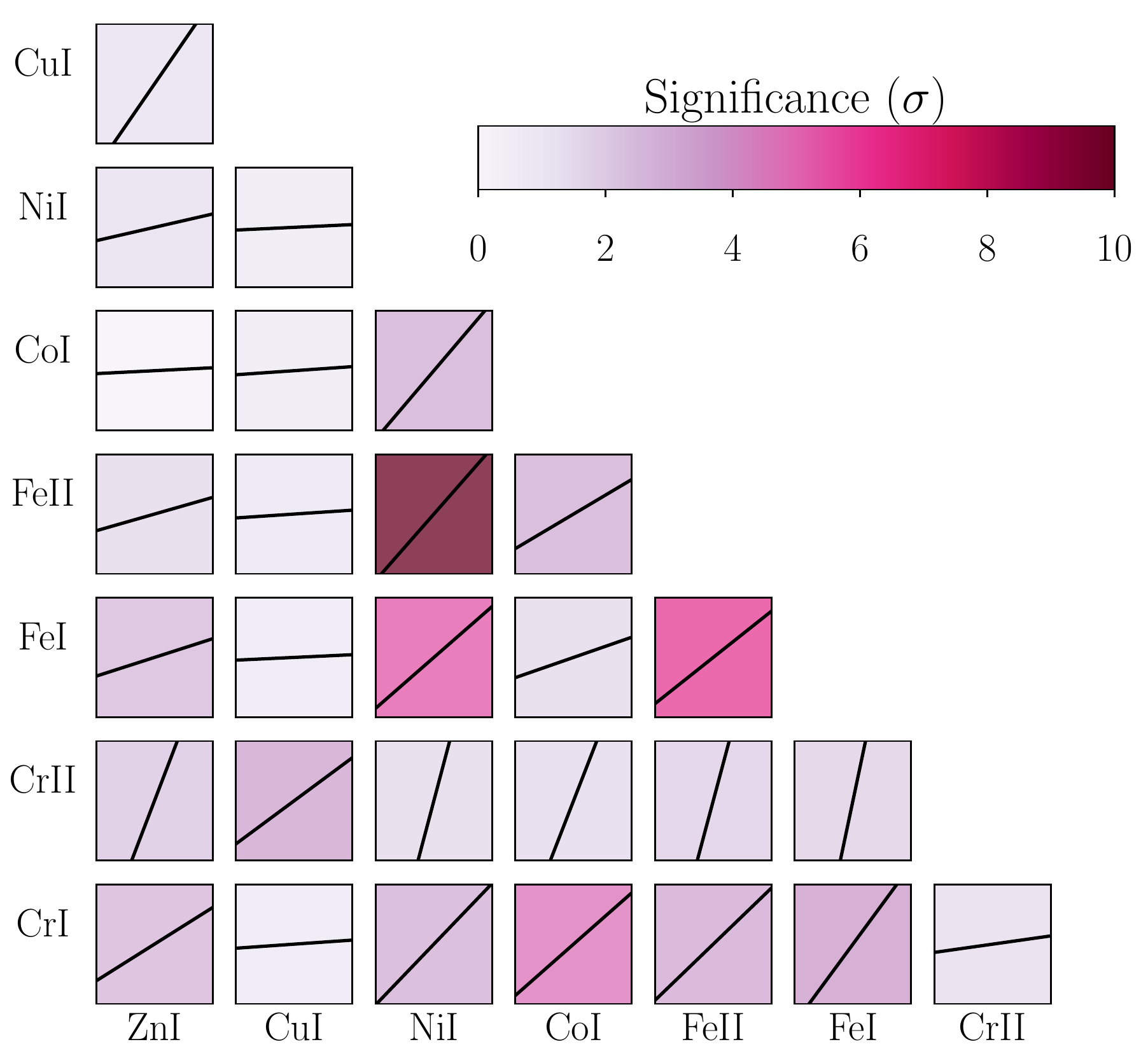}
	\caption{Fe-peak element correlations coloured by statistical significance for \textbf{NGC~362}}
    \label{fig:fepeakpyramid362}
\end{figure}

\begin{figure}
	\includegraphics[scale=0.4]{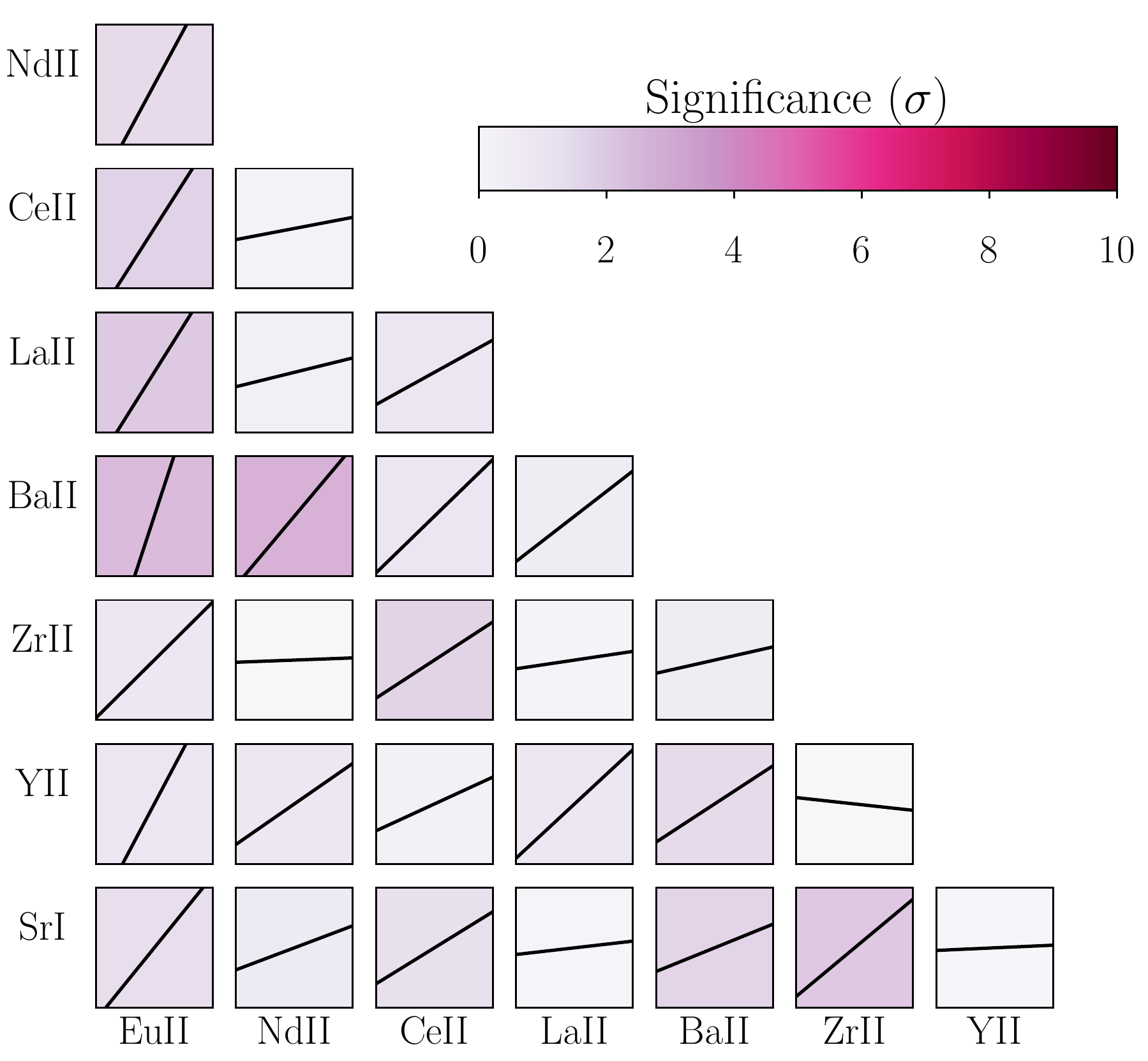}
	\caption{$s$- and $r$-process element correlations coloured by statistical significance for \textbf{NGC~288}.}
    \label{fig:sprocesspyramid288}
\end{figure}

\begin{figure}
	\includegraphics[scale=0.4]{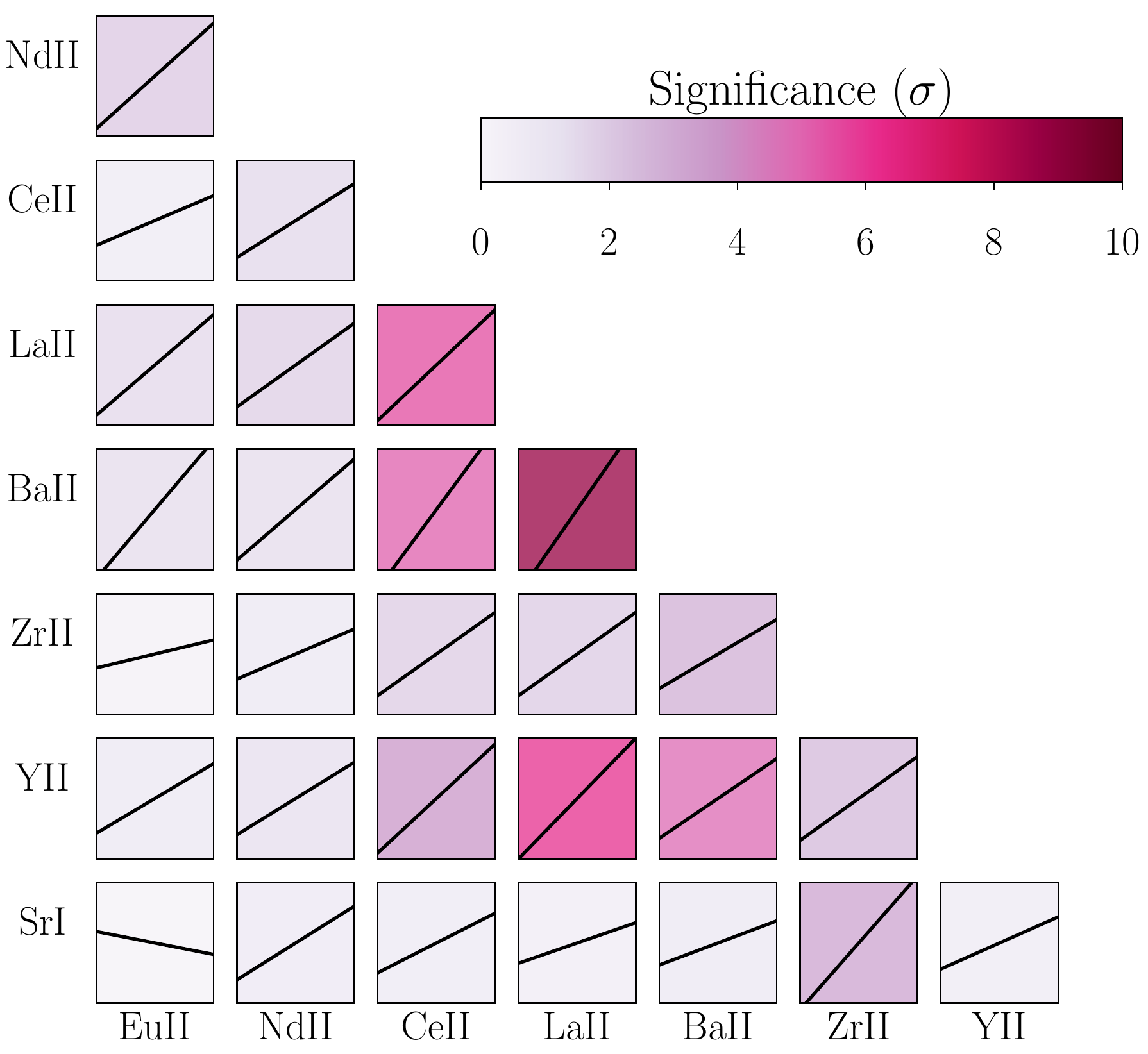}
	\caption{$s$- and $r$-process element correlations coloured by statistical significance for \textbf{NGC~362}.}
    \label{fig:sprocesspyramid362}
\end{figure}

\begin{figure*}
\includegraphics[scale=0.3]{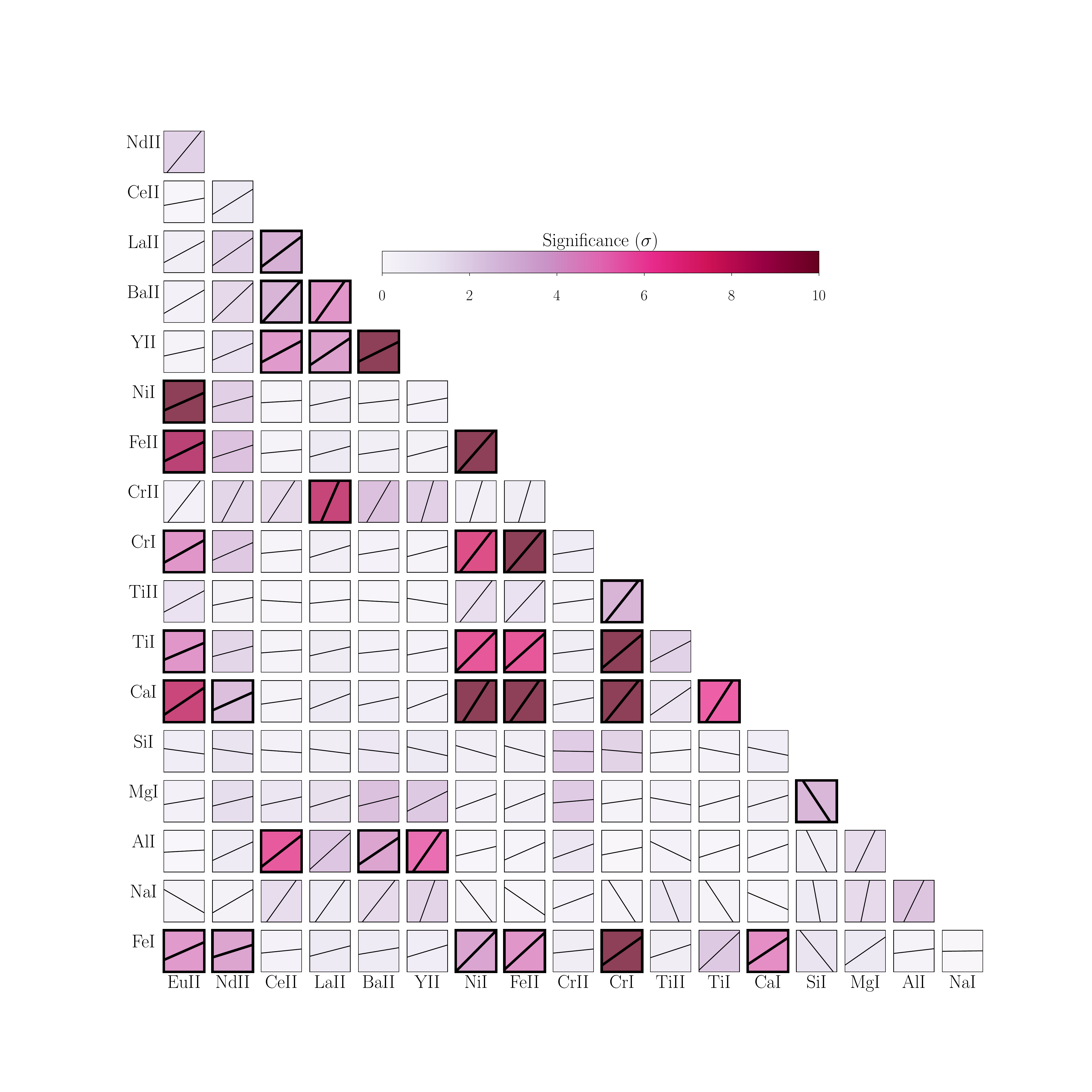}
\includegraphics[scale=0.3]{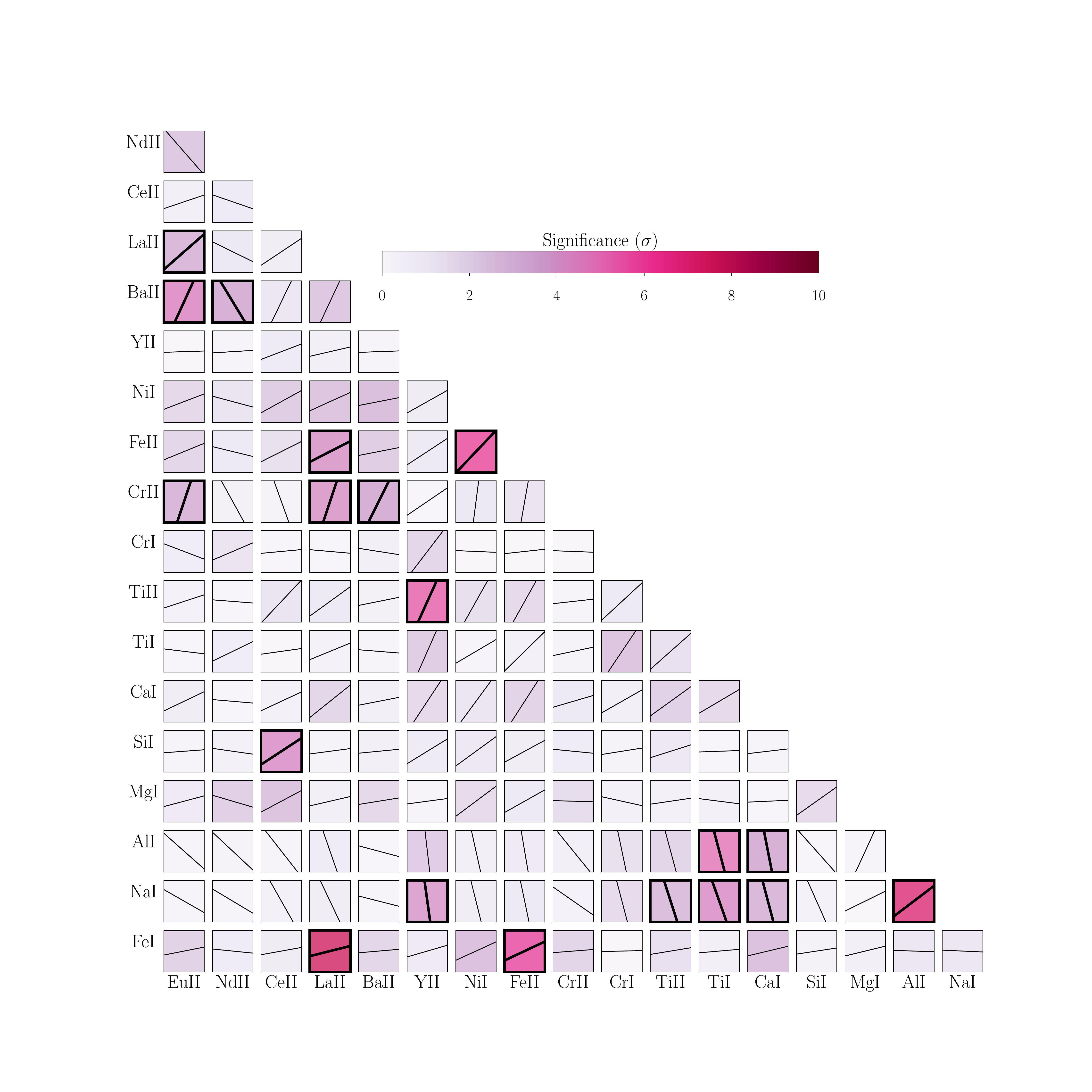}
    \caption{Same as Fig.~\ref{fig:288pyramidplot} for the two $s$-process groups in \textbf{NGC~362}. The \textbf{$s$-process rich} group is shown on the \textbf{top} and the \textbf{$s$-process weak} group on the \textbf{bottom}. Note that only in the $s$-process rich group are the $s$-process elements Ce, Ba and Y found to correlate with the light element Al - indicating enrichment via AGB stars as a natural consequence of GC evolution. The lack of correlation between the $s$-process weak group and the light element Al supports primordial enrichment in $s$-process elements within the proto-cluster environment.}
    \label{fig:362pyramidplot_sproc}
\end{figure*}

\todo{Element correlations for each nucleosynthetic group as discussed in Sec.~\ref{sec:elemcor_grouped} are shown for the $\alpha$-elements (both clusters: Fig.~\ref{fig:alphapyramid}), the Fe-peak elements (NGC~288: Fig.~\ref{fig:fepeakpyramid288}, NGC~362: Fig.~\ref{fig:fepeakpyramid362}) and heavy elements (NGC~288: Fig.~\ref{fig:sprocesspyramid288}, NGC~362: Fig.~\ref{fig:sprocesspyramid362}).} Chemical correlations for all possible element correlations in the two $s$-process groups in NGC~362 are shown in Fig.~\ref{fig:362pyramidplot_sproc}. the $s$-rich group is shown in the topmost panel and $s$-weak group below. Interpretations of the interesting and unexpected correlations are discussed in Section \ref{sec:chemcorr} and subsequent subsections.

\section{Equivalent Widths \& Stellar Abundances}
Table~\ref{tab:ews} lists a sample of equivalent width measurements for all stars in this study, including the reference star mg9. A description of how the lines are measured and which lines are included for the final analysis is given in Sec.~\ref{sec:ews}. A sample of abundances measurements for select stars in NGC~288 and NGC~362 are given in Table~\ref{tab:ngc288_abund} and Table~\ref{tab:ngc362_abund} respectively. Full tables are included in the online material.


\newpage
\begin{table*}
\caption{Sample of the equivalent widths measured for the stars in this study and used in the determination of differential abundances. Lines for the reference star, mg9 are also included. A description of the measurement methodology and choice of lines to include is given in Sec.~\ref{sec:ews}.}
\label{tab:ews}
\begin{tabular}{llllllll}
\hline
Wavelength & Element & $\chi$ & log gf & mg9 & NGC288-281 & NGC288-287 & NGC288-338 \\
{[\AA]} & ... & [eV] & ... & {[m\AA]} & {[m\AA]} & {[m\AA]} & {[m\AA]} \\
\hline
6363.78 & 8.0 & 0.02 & -10.3 & 11.1 & 26.6 & 10.3 & 19.3 \\
5682.63 & 11.0 & 2.1 & -0.71 & 56.6 & 64.9 & 80.9 & 56.1 \\
5688.2 & 11.0 & 2.1 & -0.41 & ...  & 90.7 & ...  & ...  \\
6154.226 & 11.0 & 2.1 & -1.547 & 13.7 & 16.3 & 27.1 & 13.5 \\
6160.747 & 11.0 & 2.1 & -1.246 & 25.6 & 32.4 & 43.0 & 24.1 \\
5528.4 & 12.0 & 4.35 & -0.5 & 170.6 & 197.3 & 170.1 & 195.3 \\
5711.09 & 12.0 & 4.34 & -1.72 & 84.2 & 113.5 & 98.9 & 98.3 \\
6698.673 & 13.0 & 3.14 & -1.647 & 17.2 & 19.6 & 20.2 & 14.3 \\
5645.613 & 14.0 & 4.93 & -2.14 & 14.8 & ...  & 23.1 & 22.0 \\
5665.55 & 14.0 & 4.92 & -2.04 & 16.6 & 27.0 & ...  & 25.6 \\
5684.48 & 14.0 & 4.95 & -1.42 & ...  & ...  & 32.1 & ...  \\
5690.43 & 14.0 & 4.93 & -1.87 & 22.3 & 34.3 & 32.5 & 36.4 \\
5701.1 & 14.0 & 4.93 & -2.05 & 16.9 & 26.9 & 24.9 & 26.6 \\
5948.55 & 14.0 & 5.08 & -1.23 & 47.5 & 60.9 & 65.8 & 61.5 \\
6142.49 & 14.0 & 5.62 & -1.48 & 9.6 & ...  & 14.9 & ...  \\
6155.14 & 14.0 & 5.62 & -0.86 & 29.7 & 40.0 & 42.0 & 43.7 \\
6237.33 & 14.0 & 5.62 & -1.08 & 20.3 & 29.6 & 31.2 & 29.1 \\
6243.814 & 14.0 & 5.62 & -1.244 & 13.1 & ...  & 20.7 & 22.8 \\
6244.465 & 14.0 & 5.62 & -1.091 & ...  & ...  & 21.1 & ...  \\
6721.84 & 14.0 & 5.86 & -0.94 & 14.2 & ...  & 20.8 & 18.8 \\
... & ... & ... & ... & ... & ... & ... & ... \\
\hline
\end{tabular}
\end{table*}

\begin{table*}
\caption{Sample of the stellar abundances for the GC NGC~288 determined following the methodology described in Sec.~\ref{sec:difabund}. All abundances listed are quoted in a differential sense relative to the reference star mg9. The abundances listed for mg9 are absolute abundances.}
\label{tab:ngc288_abund}
\begin{tabular}{rrrrrrrrrrrrrr}
\hline
Element & $N$ & mg9 & $\sigma$ & $N$ & NGC288-281 & $\sigma$ & $N$ & NGC288-287 & $\sigma$ & $N$ & NGC288-338 & $\sigma$ \\
\hline
\ion{Na}{I} & 3 & 4.895 & 0.025 & 4 & -0.034 & 0.015 & 3 & 0.358 & 0.018 & 3 & -0.036 & 0.016 \\
\ion{Mg}{I} & 2 & 6.351 & 0.014 & 2 & 0.279 & 0.142 & 2 & 0.134 & 0.139 & 2 & 0.225 & 0.036 \\
\ion{Al}{I} & 1 & 5.303 & ... & 1 & -0.068 & 0.007 & 1 & 0.097 & 0.011 & 1 & -0.105 & 0.013 \\
\ion{Si}{I} & 10 & 6.230 & 0.036 & 6 & 0.335 & 0.009 & 11 & 0.298 & 0.013 & 9 & 0.318 & 0.019 \\
\ion{Ca}{I} & 12 & 5.037 & 0.036 & 3 & 0.248 & 0.013 & 7 & 0.226 & 0.021 & 6 & 0.269 & 0.027 \\
\ion{Sc}{II} & 1 & 1.408 & ... & 1 & 0.440 & 0.019 & 1 & 0.446 & 0.021 & 1 & 0.413 & 0.026 \\
\ion{Ti}{I} & 39 & 3.550 & 0.068 & 18 & 0.309 & 0.017 & 33 & 0.198 & 0.026 & 23 & 0.286 & 0.028 \\
\ion{Ti}{II} & 10 & 3.574 & 0.051 & 5 & 0.261 & 0.016 & 10 & 0.179 & 0.027 & 7 & 0.233 & 0.026 \\
\ion{V}{I} & 1 & 2.339 & ... & 0 & ... & ... & 1 & 0.279 & 0.030 & 1 & 0.403 & 0.038 \\
\ion{Cr}{I} & 4 & 3.915 & 0.054 & 3 & 0.221 & 0.035 & 4 & 0.200 & 0.028 & 3 & 0.239 & 0.031 \\
\ion{Cr}{II} & 2 & 4.127 & 0.003 & 2 & 0.301 & 0.132 & 2 & 0.155 & 0.030 & 2 & 0.211 & 0.080 \\
\ion{Mn}{I} & 3 & 3.503 & 0.029 & 0 & ... & ... & 2 & 0.374 & 0.131 & 2 & 0.445 & 0.234 \\
\ion{Fe}{I} & 130 & 5.849 & 0.093 & 97 & 0.250 & 0.008 & 105 & 0.191 & 0.013 & 105 & 0.231 & 0.015 \\
\ion{Fe}{II} & 15 & 5.770 & 0.032 & 16 & 0.263 & 0.026 & 14 & 0.213 & 0.027 & 13 & 0.251 & 0.033 \\
\ion{Co}{I} & 2 & 3.504 & 0.236 & 1 & 0.279 & 0.009 & 2 & 0.194 & 0.018 & 1 & 0.285 & 0.015 \\
\ion{Ni}{I} & 43 & 4.524 & 0.107 & 23 & 0.298 & 0.011 & 38 & 0.237 & 0.018 & 31 & 0.291 & 0.014 \\
\ion{Cu}{I} & 1 & 2.416 & ... & 1 & 0.844 & 0.025 & 1 & 0.383 & 0.027 & 1 & 0.653 & 0.035 \\
\ion{Zn}{I} & 1 & 2.951 & ... & 1 & -0.051 & 0.010 & 1 & 0.179 & 0.016 & 1 & 0.144 & 0.019 \\
\ion{Sr}{I} & 1 & 3.548 & ... & 1 & 0.111 & 0.012 & 1 & 0.102 & 0.016 & 1 & 0.184 & 0.020 \\
\ion{Y}{II} & 8 & 0.582 & 0.096 & 5 & 0.467 & 0.047 & 7 & 0.296 & 0.030 & 7 & 0.402 & 0.031 \\
\ion{Zr}{II} & 1 & 1.446 & ... & 1 & 0.325 & 0.018 & 1 & 0.355 & 0.018 & 1 & 0.420 & 0.023 \\
\ion{Ba}{II} & 3 & -0.175 & 0.049 & 3 & 0.343 & 0.042 & 3 & 0.194 & 0.039 & 3 & 0.386 & 0.035 \\
\ion{La}{II} & 3 & -0.627 & 0.074 & 2 & 0.357 & 0.058 & 2 & 0.303 & 0.035 & 2 & 0.395 & 0.029 \\
\ion{Ce}{II} & 3 & -0.080 & 0.057 & 3 & 0.376 & 0.040 & 2 & 0.340 & 0.018 & 3 & 0.444 & 0.043 \\
\ion{Nd}{II} & 11 & -0.025 & 0.044 & 4 & 0.400 & 0.020 & 11 & 0.277 & 0.023 & 8 & 0.395 & 0.030 \\
\ion{Sm}{II} & 1 & -0.438 & ... & 1 & 0.523 & 0.016 & 1 & 0.358 & 0.017 & 0 & ... & ... \\
\ion{Eu}{II} & 1 & -1.240 & ... & 1 & 0.401 & 0.020 & 1 & 0.370 & 0.020 & 1 & 0.433 & 0.025 \\
\hline
\end{tabular}
\end{table*}

\begin{table*}
\caption{Same as Table~\ref{tab:ngc288_abund}, for the GC NGC~362.}
\label{tab:ngc362_abund}
\begin{tabular}{rrrrrrrrrrrrr}
\hline
Element & $N$ & mg9 & $\sigma$ & $N$  & NGC362-1137 & $\sigma$  & $N$  & NGC362-1334 & $\sigma$  & $N$  & NGC362-2127 & $\sigma$  \\
\hline
\ion{Na}{I} & 3 & 4.895 & 0.025 & 4 & -0.074 & 0.033 & 3 & -0.022 & 0.035 & 3 & 0.361 & 0.028 \\
\ion{Mg}{I} & 2 & 6.351 & 0.014 & 2 & 0.270 & 0.088 & 2 & 0.241 & 0.144 & 2 & 0.275 & 0.050 \\
\ion{Al}{I} & 1 & 5.303 & ... & 1 & -0.329 & 0.018 & 1 & -0.181 & 0.021 & 1 & 0.200 & 0.013 \\
\ion{Si}{I} & 10 & 6.230 & 0.036 & 7 & 0.286 & 0.023 & 7 & 0.324 & 0.027 & 8 & 0.255 & 0.018 \\
\ion{Ca}{I} & 12 & 5.037 & 0.036 & 4 & 0.295 & 0.030 & 3 & 0.319 & 0.032 & 5 & 0.216 & 0.024 \\
\ion{Sc}{II} & 1 & 1.408 & ... & 1 & 0.508 & 0.037 & 1 & 0.546 & 0.038 & 1 & 0.401 & 0.029 \\
\ion{Ti}{I} & 39 & 3.550 & 0.068 & 18 & 0.368 & 0.039 & 14 & 0.377 & 0.046 & 23 & 0.218 & 0.029 \\
\ion{Ti}{II} & 10 & 3.574 & 0.051 & 4 & 0.297 & 0.035 & 4 & 0.363 & 0.037 & 5 & 0.175 & 0.026 \\
\ion{V}{I} & 1 & 2.339 & ... & 1 & 0.822 & 0.066 & 1 & 0.946 & 0.085 & 1 & 0.453 & 0.040 \\
\ion{Cr}{I} & 4 & 3.915 & 0.054 & 2 & 0.305 & 0.042 & 2 & 0.348 & 0.043 & 3 & 0.209 & 0.052 \\
\ion{Cr}{II} & 2 & 4.127 & 0.003 & 2 & 0.472 & 0.097 & 1 & 0.392 & 0.038 & 1 & 0.457 & 0.029 \\
\ion{Mn}{I} & 3 & 3.503 & 0.029 & 3 & 1.314 & 0.249 & 3 & 1.367 & 0.185 & 3 & 0.884 & 0.124 \\
\ion{Fe}{I} & 130 & 5.849 & 0.093 & 77 & 0.371 & 0.017 & 74 & 0.421 & 0.020 & 72 & 0.359 & 0.012 \\
\ion{Fe}{II} & 15 & 5.770 & 0.032 & 14 & 0.366 & 0.055 & 13 & 0.421 & 0.059 & 15 & 0.341 & 0.042 \\
\ion{Co}{I} & 2 & 3.504 & 0.236 & 1 & 0.363 & 0.017 & 1 & 0.404 & 0.019 & 1 & 0.310 & 0.013 \\
\ion{Ni}{I} & 43 & 4.524 & 0.107 & 24 & 0.325 & 0.020 & 24 & 0.384 & 0.024 & 27 & 0.307 & 0.017 \\
\ion{Cu}{I} & 1 & 2.416 & ... & 1 & 0.796 & 0.065 & 1 & 0.299 & 0.066 & 1 & 0.576 & 0.048 \\
\ion{Zn}{I} & 1 & 2.951 & ... & 0 & ... & ... & 1 & 0.252 & 0.038 & 1 & 0.124 & 0.025 \\
\ion{Sr}{I} & 1 & 3.548 & ... & 1 & 0.051 & 0.030 & 1 & 0.073 & 0.034 & 0 & ... & ... \\
\ion{Y}{II} & 8 & 0.582 & 0.096 & 3 & 0.309 & 0.030 & 2 & 0.501 & 0.031 & 3 & 0.247 & 0.053 \\
\ion{Zr}{II} & 1 & 1.446 & ... & 1 & 0.269 & 0.028 & 1 & 0.388 & 0.029 & 1 & 0.281 & 0.022 \\
\ion{Ba}{II} & 3 & 0.099 & 0.396 & 2 & 0.440 & 0.041 & 2 & 0.573 & 0.048 & 2 & 0.479 & 0.074 \\
\ion{La}{II} & 3 & -0.627 & 0.074 & 3 & 0.426 & 0.032 & 3 & 0.530 & 0.045 & 3 & 0.404 & 0.042 \\
\ion{Ce}{II} & 3 & -0.080 & 0.057 & 2 & 0.376 & 0.033 & 2 & 0.339 & 0.080 & 2 & 0.338 & 0.031 \\
\ion{Nd}{II} & 11 & -0.025 & 0.044 & 5 & 0.538 & 0.039 & 5 & 0.688 & 0.065 & 6 & 0.463 & 0.038 \\
\ion{Sm}{II} & 1 & -0.438 & ... & 0 & ... & ... & 0 & ... & ... & 0 & ... & ... \\
\ion{Eu}{II} & 1 & -1.240 & ... & 1 & 0.563 & 0.032 & 1 & 0.725 & 0.035 & 1 & 0.592 & 0.027 \\
\hline
\end{tabular}
\end{table*}

\bsp	
\label{lastpage}
\end{document}